\documentclass[amsmath,amssymb,aps,prc,reprint,superscriptaddress]{revtex4-2}
\usepackage{booktabs}
\usepackage{graphicx}
\usepackage{float}
\usepackage{xcolor}
\usepackage{dcolumn}

\usepackage{bm}
\usepackage{hyperref}

\usepackage[margin=0.8in]{geometry}

\usepackage{siunitx}
\usepackage{array}
\usepackage{longtable} 
\usepackage{tabularx}  
\usepackage{geometry}

\begin{document}

\title{Neutron-induced reaction cross section measurements on carbon at neutron energies up to 55 MeV at LANSCE}

\author{A. Wantz}
  \email{wantz@frib.msu.edu}
\affiliation{Facility for Rare Isotope Beams, Michigan State University, East Lansing, Michigan, 48824}
\affiliation{Physics and Astronomy Department, Michigan State University, East Lansing, Michigan, 48824}

\author{A. N. Kuchera}
\affiliation{Department of Physics, Davidson College, Davidson, North Carolina, 28305}

\author{S. A. Kuvin}%
\affiliation{Los Alamos National Laboratory, Los Alamos, New Mexico, 87545}

\author{W. F. Rogers}%
\affiliation{Indiana Wesleyan University, Marion, Indiana, 46953}

\author{T. Baumann}
\affiliation{Facility for Rare Isotope Beams, Michigan State University, East Lansing, Michigan, 48824}

\author{M. Devlin}%
\affiliation{Los Alamos National Laboratory, Los Alamos, New Mexico, 87545}

\author{N. Frank}
\affiliation{Augustana College, Rock Island, Illinois, 61201}

\author{P. Gueye}
\affiliation{Facility for Rare Isotope Beams, Michigan State University, East Lansing, Michigan, 48824}
\affiliation{Physics and Astronomy Department, Michigan State University, East Lansing, Michigan, 48824}

\author{K. J. Kelly}
\affiliation{Los Alamos National Laboratory, Los Alamos, New Mexico, 87545}

\author{T. Redpath}
\affiliation{Virgina State University, Petersburg, Virginia, 23806}

\author{J. R. Winkelbauer}
\affiliation{Los Alamos National Laboratory, Los Alamos, New Mexico, 87545}

\author{A. LaRochelle}
\affiliation{Department of Physics, Davidson College, Davidson, North Carolina, 28305}

\author{R. Devlin}
\affiliation{Department of Physics, Davidson College, Davidson, North Carolina, 28305}

\author{D. Flores-Madrid}
\affiliation{Indiana Wesleyan University, Marion, Indiana, 46953}

\author{B. Hassan}
\affiliation{Department of Physics, Davidson College, Davidson, North Carolina, 28305}

\author{O. Lucas}
\affiliation{Indiana Wesleyan University, Marion, Indiana, 46953}

\author{A. Maki}
\affiliation{Department of Physics, Davidson College, Davidson, North Carolina, 28305}

\author{N. Mendez}
\affiliation{Facility for Rare Isotope Beams, Michigan State University, East Lansing, Michigan, 48824}
\affiliation{Physics and Astronomy Department, Michigan State University, East Lansing, Michigan, 48824}
\affiliation{Los Alamos National Laboratory, Los Alamos, New Mexico, 87545}

\author{T. Sandy}
\affiliation{Department of Physics, Davidson College, Davidson, North Carolina, 28305}

\author{J. Smith}
\affiliation{Taylor University, Upland, Indiana, 46989}

\author{W. Sussenbach}
\affiliation{Indiana Wesleyan University, Marion, Indiana, 46953}

\author{G. Votta}
\affiliation{Facility for Rare Isotope Beams, Michigan State University, East Lansing, Michigan, 48824}
\affiliation{Physics and Astronomy Department, Michigan State University, East Lansing, Michigan, 48824}

\author{K. Wang}
\affiliation{Department of Physics, Davidson College, Davidson, North Carolina, 28305}

\author{S. Winner}
\affiliation{Department of Physics, Davidson College, Davidson, North Carolina, 28305}

\date{\today}

\begin{abstract}
\textbf{Background:}  Single-crystal chemical vapor deposited (sCVD) diamond detectors offer a unique method to study cross sections of reactions on carbon since they can be used as active targets. Previous studies analyzing neutrons on carbon using these detectors were primarily focused on lower energy neutrons and reactions, and some of these did not have sufficient energy resolution to isolate the contributions of reaction channels with similar Q values. 

\textbf{Purpose:} This work extends neutron-induced reaction cross section measurements to higher energies, relevant to rare isotope facilities. These measurements can be used to inform and benchmark simulation of experiments that require neutron detection, particularly those utilizing organic scintillators. For some experiments, simulations are used to extract physics information from experimental data, reinforcing the need for accurate simulations.

\textbf{Methods:} Two sCVD diamond detectors were used as active targets at LANSCE, where neutrons up to 800 MeV are produced via spallation. 

\textbf{Results:} Relative cross sections are reported from incident neutron (kinetic) energies E$_n$ = 12 MeV up to 55 MeV for $^{12}$C(n,$\alpha_0$), up to 46 MeV for $^{12}$C(n,d$_0$), and up to 27 MeV for $^{12}$C(n,p$_0$) and $^{12}$C(n,p$_1$). These measurements extend these cross sections to higher energies than those of previous studies.

\textbf{Conclusions:} Good agreement is found between this work and recent experimental data from the EXFOR database in the neutron energies where the studies overlap. This work supports the need to update the ENDF evaluation for the (n,$\alpha_0$) channel with more recent data, and provides data that could allow for an evaluation of the (n,p$_0$), (n,p$_1$), and (n,d$_0$) channels. These cross sections will increase the accuracy of simulations by extending the energy range for which empirical cross sections are available and including new reaction channels.

\end{abstract}

\maketitle

\section{\label{sec:level1}Introduction }
Neutron detectors are crucial for numerous areas of physics, including neutrino experiments such as DUNE\cite{Abi_2020}, NO$\nu$A \cite{NOvA:2007rmc}, T2K \cite{ABE2011106}, and MINER$\nu$A \cite{neutrino} and nuclear structure experiments with neutron-rich nuclei. In particular, large-volume organic scintillator arrays such as MoNA \cite{LUTHER200333}, NEBULA \cite{NAKAMURA2016156}, and NeuLAND \cite{BORETZKY2021165701} allow for high detection efficiency for fast neutrons. The Facility for Rare Isotope Beams (FRIB) will expand access to neutron-rich nuclei, necessitating accurate neutron detection and simulation for studies of beta-delayed neutron emitters and neutron-unbound systems \cite{NSACLRP}. One approach to simulating these experiments and detectors is to use MENATE$\_$R  \cite{KOHLEY201259}. MENATE$\_$R uses measured n-H and n-C interaction cross section data to calculate the likelihood of each possible interaction, determine which, if any, reaction takes place, and then simulates the interaction within GEANT4 \cite{AGOSTINELLI2003250, ALLISON2016186, GEANT4_2nd_reference}. 

A previous study \cite{ROGERS2019} compared MENATE$\_$R simulations  with experimental data taken at the Los Alamos Neutron Science Center (LANSCE) and with simulations utilizing the GEANT4 physics lists. The latter do not use experimental data above 20 MeV, instead utilizing a cascade model. Some observables showed better agreement between data and one or the other simulation package in specific energy regimes; hence, a universal recommendation for the preferred package could not be made. This helped to illustrate that MENATE$\_$R needed improvement, particularly partial cross section measurements of reactions on carbon, for new reaction channels and at higher energies. Existing data are sparse or nonexistent at higher energies, and more data are needed to accurately characterize these processes. 
\begin{figure*}[ht]
\includegraphics[width=1.8\columnwidth]{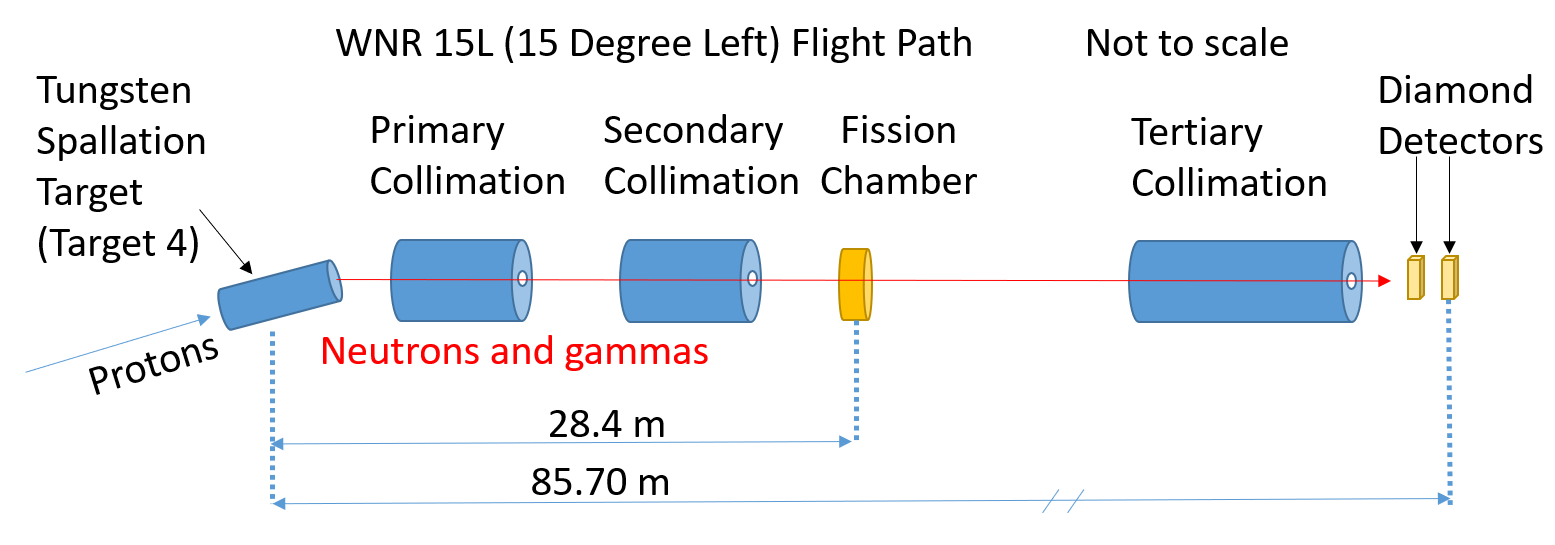}
\caption{\label{fig:Setup} A diagram of the 15-degree left flight path at the WNR facility within LANSCE. Two Cividec B8 diamond detectors were placed in the 90 m station, downstream of the tertiary collimation. Only the first 6.54 m of the flight path is under vacuum. The diagram is not drawn to scale.}
\end{figure*}

Numerous studies have demonstrated the capability of diamond detectors to determine neutron-induced reaction cross sections on carbon \cite{Kuvin2021, MAJERLE2020163014, Pillon2011, Pillon2017},  due to their good energy resolution and high count rate capabilities.  The present study builds on those works with a unique combination of improvements: using a white neutron source to gain access to a continuous distribution of incident neutron energies, including higher incident neutron energies than have been analyzed previously, a longer flight path to improve the resolving power by spacing out the arrival time of neutrons of different energies, and using a spectroscopic amplifier with improved energy resolution to isolate contributions of reaction channels with similar Q values. 

This study highlights four different $^{12}$C(n,X) reaction channels, presenting higher energy and/or higher resolution energy-differential partial cross section data than has been measured previously, shedding light on the reaction channels that MENATE$\_$R currently lacks. Additionally, this work provides insight into the likelihood of dark scattering: in a neutron-carbon reaction, the recoiling ionizing particle may or may not generate a measurable signal above threshold in the scintillator. Such events that scatter on carbon but do not trigger the detector are referred to as ``dark scatter," and contribute to uncertainties in neutron event reconstruction in large detector arrays like MoNA. By improving cross section data for individual reaction channels, the portion of the total cross section in scintillator that is dark scatter is better constrained. Furthermore, this work supports ongoing efforts to update the ENDF/B-VIII.0 library \cite{ENDF} with new data to resolve consistent systematic differences between the older cross section measurements on which ENDF is based and more recent experimental results.

\section{Experiment}
The proton accelerator at LANSCE sent 800 MeV protons to the spallation target at the Weapons Neutron Research facility (WNR, Target 4),  LANSCE uses a pulsed proton beam structure, which consists of macropulses and micropulses. 120 macropulses are produced per second, of which 100 are sent to WNR (Target 4), and the remaining 20 are sent to the nearby Lujan Center. Each macropulse is 625 $\mu$s in duration, and macropulses are broken up into micropulses that are separated by 1.8 $\mu$s. Thus, there are 1.8 $\mu$s separating each group of neutrons. 

Two Cividec B8 diamond detectors \cite{Cividec} were placed in the 90 m station on the 15-degree flight path (4FP15L). The beam was collimated to a 3 mm diameter at the entrance into the station, slightly smaller than the 4x4 mm$^2$ active area of the diamonds,  and the diamonds were placed near the entrance collimator. A diagram of the setup is included in Fig. \ref{fig:Setup}. The diamond detectors are the same model, but their signals were processed differently. The downstream diamond detector was coupled with the Cividec C6 fast charge amplifier, while the upstream diamond detector utilized the Cividec Cx spectroscopic shaping amplifier. Due to the high gain of the Cx amplifier, the signal output from the upstream detector was attenuated by a factor of 10. The signals were routed into a CAEN VX1730 digitizer (14 bit, 500 MS/s) with pulse shape discrimination (PSD) firmware. The rise times for the detector signals were 8 ns and 80 ns for the detectors with the C6 and Cx amplifiers, respectively. Waveforms were not recorded, only the waveform integrals, due to data storage constraints. This limited the amount of analysis that could be done with the diamond operated with the C6 amplifier, which benefits from offline analysis of the waveforms. However, even if waveforms had been recorded, the resolution using the Cx amplifier would still have been superior. Information about the diamond detector setup is included in Table \ref{tab:DetInfoTable}.

\begin{table}[h]
\caption{\label{tab:DetInfoTable}
Summary of the detector characteristics and run details for the experiment.}
\begin{ruledtabular}
\begin{tabular}{lcc}
Data collection time &\multicolumn{2}{c}{253 hours} \\
Diamond detector thickness &\multicolumn{2}{c}{0.5 mm  each}  \\
Timing resolution &\multicolumn{2}{c}{1.0 ns}  \\  \hline
\textbf{Detector} & \textbf{Upstream}  & \textbf{Downstream}  \\ \hline
Amplifier  & Cx & C6 \\ 
Attenuation setting & 0.1 & 1.0 \\
Detector bias & +400 V & +500 V \\
Distance from Target 4 & 85.69 m  &  85.70 m \\
\end{tabular}
\end{ruledtabular}
\end{table} 
Because of the relatively long distance from the spallation target to the diamond detectors, the slowest neutrons of one pulse arrive later than the fastest neutrons of the next pulse, creating a background known as wraparound neutrons. For this flight path distance (85.7 m), the wraparound neutrons are at 12 MeV at the start of the next pulse. Because of the wraparound neutrons, and because of the higher energy focus of this project, the analysis was restricted to incident neutron energies above 12 MeV. The wraparound neutron background is easily excluded from the events of interest above E$_n$ = 12 MeV since they do not overlap with the reconstructed Q values of the reactions studied in this work. Thus, the wraparound neutrons did not affect the analysis beyond setting a cutoff for the lowest energy neutrons that were analyzed. The average trigger rate for the detectors was a few hundred counts per second, significantly below the rate limitations of the detector and acquisition system, resulting in negligible dead time. The proton current from the accelerator ranged from 0-3.5 $\mu$A, with most of the data between 2 and 3 $\mu$A. 

\begin{figure}[tb]
\includegraphics[width=\columnwidth]{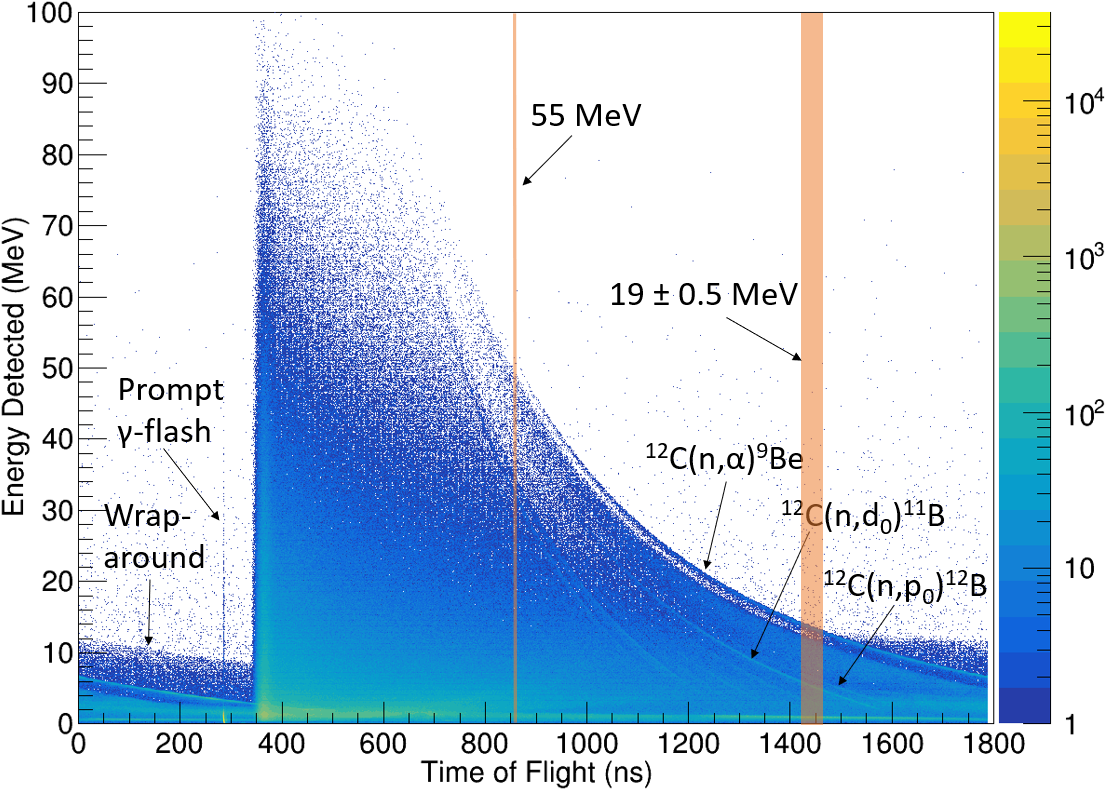}
\caption{\label{fig:energyvstof} Energy detected in diamond detector as a function of time of flight from the spallation target to the diamond detector. Notable features include the $\gamma$ flash, the wraparound neutron background, and different reaction channels. The highest energy at which cross sections were measured in this study, 55 MeV, is labeled as well. }
\end{figure}

 ``T0" timing signals, which originate from the detection of protons just before the spallation target, were routed into the data acquisition system. The T0 time is subtracted from the time at which a neutron is detected in our detectors, providing the neutron time of flight (ToF). The ToF is calibrated using the arrival time of the prompt gamma rays from the spallation target and the known energy of the fastest-moving neutrons (800 MeV). From this ToF, the incident neutron energy is calculated. The energy detected in the diamond is plotted against the time of flight in Fig. \ref{fig:energyvstof}.

\begin{figure*}
\includegraphics[width=1.8\columnwidth]{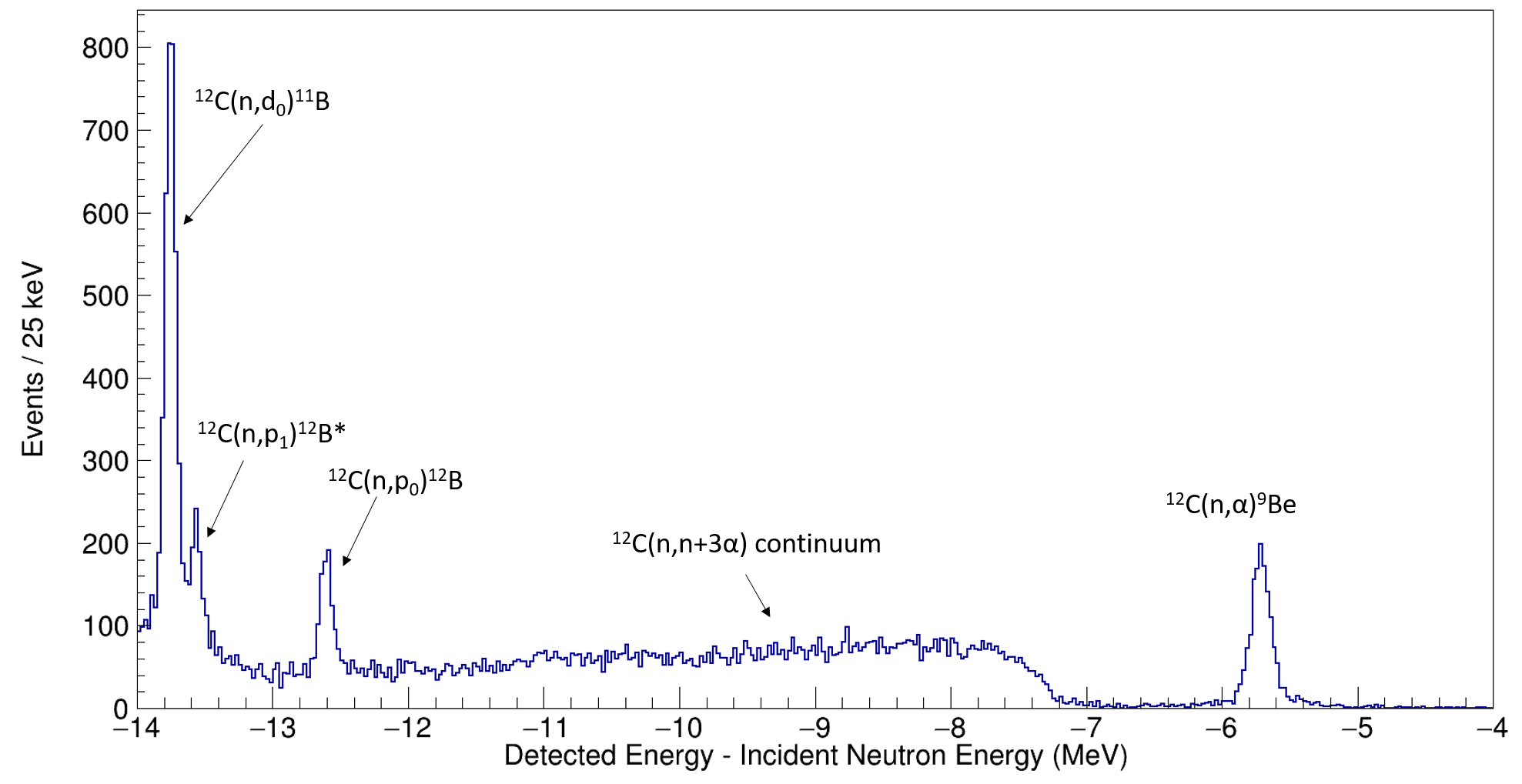}
\caption{\label{fig:qval} Reconstructed reaction Q-value spectrum for neutrons with E$_n$ = 19 $\pm$ 0.5 MeV, from the higher resolution (upstream) detector. This allowed for identification of the peaks corresponding to different reaction channels. For context, the data in this spectrum correspond to the shaded region around 1450 ns in  Fig. \ref{fig:energyvstof}. }
\end{figure*}

The data are more easily visualized by plotting the energy detected in the diamond detector minus the incident neutron energy. This calculation reconstructs the reaction Q value for each event, in the case of complete energy deposition of the reaction products. In Fig. \ref{fig:qval}, the data is shown for the upstream diamond detector using the shaping amplifier, for 19 $\pm$ 0.5 MeV  neutrons. In this energy range, four sharp peaks are present, corresponding to reactions with only charged particles in the exit channel. These peaks were identified by their Q values, and labeled accordingly. The subscript refers to the state of the daughter nucleus that is populated, with 0 for the ground state and 1 for the first excited state. Other reaction channels and higher excited states are populated as one moves to higher incident neutron energy, once the thresholds for these reactions have been reached. 

\section{Normalization}
To determine the incoming neutron flux, a fission chamber was placed upstream of the 90 m station, at the Chi-Nu/CoGNAC  detector location, 28.4 m from the spallation target. The fission chamber consists of a $^{235}$U foil and gaseous ionization chamber. The evaluated neutron-induced fission cross sections \cite{CARLSON2018143} were used to determine the neutron flux from the number of fission fragments detected. The absolute flux in the 90 m station could not be determined with certainty, due to several factors including the difference in beam spot sizes at the fission chamber location and diamond detector location, a lack of fission chamber data for some of the runs, uncertainties in the effect of collimation on the neutron flux into the 90 m station, uncertainties in the number of target nuclei in the diamonds, and the effect of passing through an additional $\sim$60 m of air between the fission chamber and diamond detectors.
Instead, the relative flux distribution was used to compare the yields relative to the neutron flux at different incident neutron energies. The fission chamber data were compared with Monte Carlo N-Particle (MCNP) \cite{TechReport_MCNP, Taddeucci} simulations of the flux to inform the uncertainty quantification.

\subsection{Neutron flux determination}
\begin{figure}
\includegraphics[width=\columnwidth]{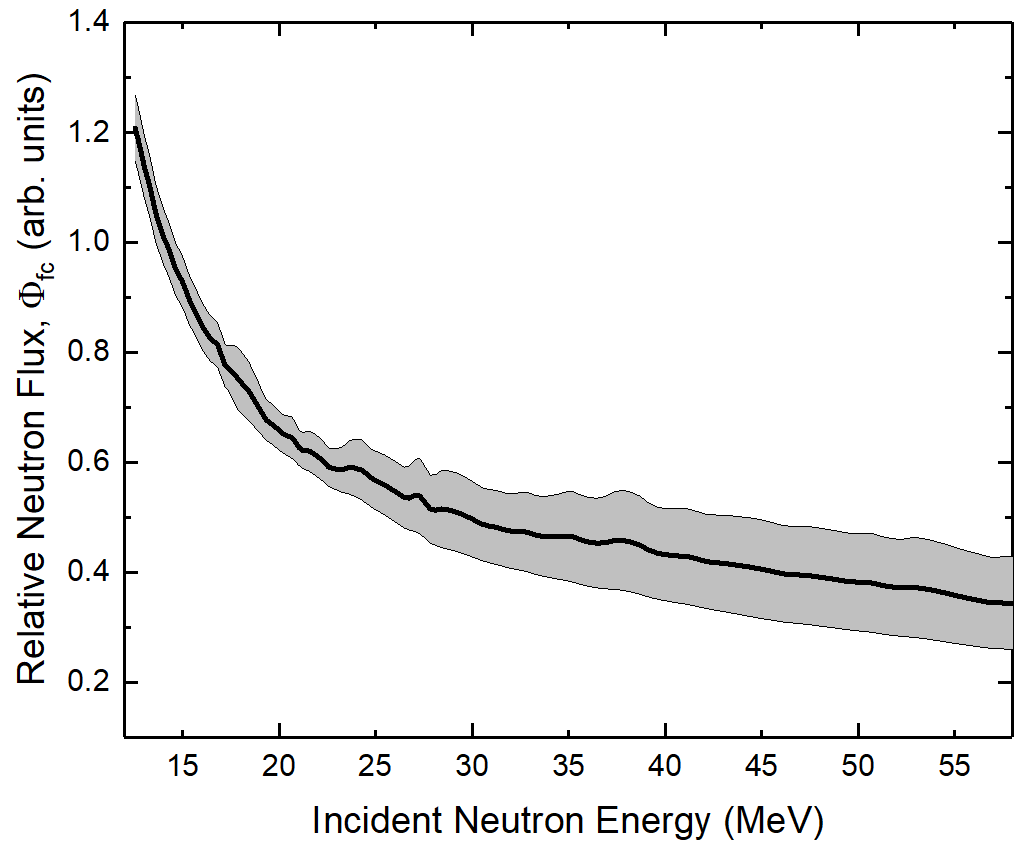}
\caption{\label{fig:Flux} The relative neutron flux distribution used for cross section normalization. The distribution was obtained with a fission chamber, using the reference standard $^{235}$U(n,f) cross sections \cite{CARLSON2018143} to calculate flux from counts. The error band shows the uncertainty in the flux, which was estimated using the differences between the flux derived from the fission chamber and MCNP simulations of the flux. The flux shown here is normalized to unity at 14.1 MeV.}
\end{figure}

The fission chamber counts were converted to flux as follows: Using the $^{235}$U(n,f) cross section ($\sigma_\text{n,f}$) reference standard \cite{CARLSON2018143}, with a linear interpolation to find the cross sections at our data bin centroids, the flux from the fission chamber data, $\Phi_\text{fc}$, is: 
\begin{equation}
    \Phi_\text{fc} = \frac{N_\text{fc}}{\sigma_\text{n,f}}
\end{equation}
where $N_\text{fc}$ is the counts in a given bin in the fission chamber data. The flux distribution used for normalization is shown in Fig. \ref{fig:Flux}. The fission chamber flux distribution was corrected to account for neutron attenuation through air. 

\subsection{Neutron attenuation simulations}
At the LANSCE facility, only a small portion of the neutron beam line is kept under vacuum. For most of the 15L flight path to the 90 m station, the neutrons pass through open air. GEANT4 simulations were performed to estimate the effect of neutrons passing through 57.3 meters of air between the fission chamber and the diamond detectors on the shape of the neutron flux distribution. The simulated neutron transmission fluctuates around 82\% from 12 MeV to 20 MeV, due to resonances in the cross sections, and then increases linearly to 89\% at 55 MeV.
 
\subsection{Cross section normalization}
Because the absolute neutron flux was not determined, the cross section measurements were normalized to previous data \cite{Pillon2011,Schmidt1992,Haight1984,Kondo2008,SANAMI2000403} from the EXFOR database, using the same method described in \cite{Kuvin2021}. The weighted average of these previous measurements, 63.6 mb ($\sigma_{\text{n,\ensuremath{\alpha}\ accepted}}$), with an uncertainty of 4.9\%, was taken as an accepted value for the $^{12}$C(n,$\alpha_0$)$^9$Be cross section at 14.1 MeV. To generate cross sections at other incident neutron energies, one can calculate:
\begin{equation}
    \sigma= \gamma \sigma_{\text{n,\ensuremath{\alpha}\ accepted}}
\end{equation}
where the normalization factor, $\gamma$, is 

\begin{equation}
  \gamma = \frac{N \Phi_0  N_\text{target} \varepsilon_0}{N_\text{n,\ensuremath{\alpha}} \Phi N_\text{target} \varepsilon }
\end{equation}
where \textit{N} represents the counts measured in a given bin for a particular reaction channel (after background subtraction), $\Phi_0$ represents the neutron flux from the fission chamber at 14.1 MeV, $\varepsilon_0$ represents the detection efficiency of the $^{12}$C(n,$\alpha_0$)$^9$Be reaction products at 14.1 MeV,  $N_\text{n,\ensuremath{\alpha}}$ represents the (n,$\alpha$) counts detected in the 14.1 MeV bin, $\Phi$ is the neutron flux obtained from the fission chamber, and $\varepsilon$ is the detection efficiency of the charged particles produced in the reaction. $N_\text{target}$, the number of target nuclei in the diamond detectors, cancels out, as long as the same detectors are used to obtain \textit{N} and $N_\text{n,\ensuremath{\alpha}}$. \textit{N}, $\varepsilon$, and $\Phi$ are energy dependent quantities, while $\Phi_0$, $\varepsilon_0$, and $N_\text{n,\ensuremath{\alpha}}$ are constant. 

\subsection{Background}
\subsubsection{External background}
The small size of the diamond detectors drastically limits the contributions of external background. The background from cosmic muons is negligible, due to the small size of the diamond detector. Event rates due to muons are five orders of magnitude less than that of the neutron beam.

Thermal neutrons have insufficient energy to trigger the detectors above threshold. Some light charged particles are produced in the spallation target, but these are swept away by a magnet and attenuated by the air between the spallation target and the diamond location. Some events detected in the diamond are caused by reactions in air via $^{14}$N(n,p)$^{14}$C, which may be responsible for some of the counts that lie above the $^{12}$C(n,$\alpha_0$)$^9$Be peak, but the quantity of these events is less than 1\% that of the $^{12}$C(n,$\alpha_0$) peak for the energies of interest.

\subsubsection{Internal background}
The primary sources of background are due to neutrons interacting in the diamond and depositing less than their full energy. This can occur in a few different ways. First, there are a number of possible continuum reactions that have one or more neutrons in the exit channel, including $^{12}$C(n,n)$^{12}$C  (elastic scattering), $^{12}$C(n,n+3$\alpha$), $^{12}$C(n,2n)$^{11}$C, and $^{12}$C(n,n+p)$^{11}$B. These outgoing neutrons carry away some of the energy, so that the Q value is not reconstructed accurately. Additionally, at higher incident neutron energies, ions from the various charged-particle-producing reactions have enough kinetic energy that they are not fully stopped in the diamond, leading to an incomplete energy-deposition tail for each of the peaks. These tails and continuum events combine to form a background for all peaks except for the $^{12}$C(n,$\alpha_0$)$^9$Be. To find the counts in each peak, the data were fit with a Gaussian plus a linear background, which reproduces the shape of the data well. The high energy of the wraparound neutron background complicates analysis below 12 MeV, and was beyond the scope of this work.  

\subsection{Charged-particle efficiency correction}
\begin{figure}
\includegraphics[width=\columnwidth]{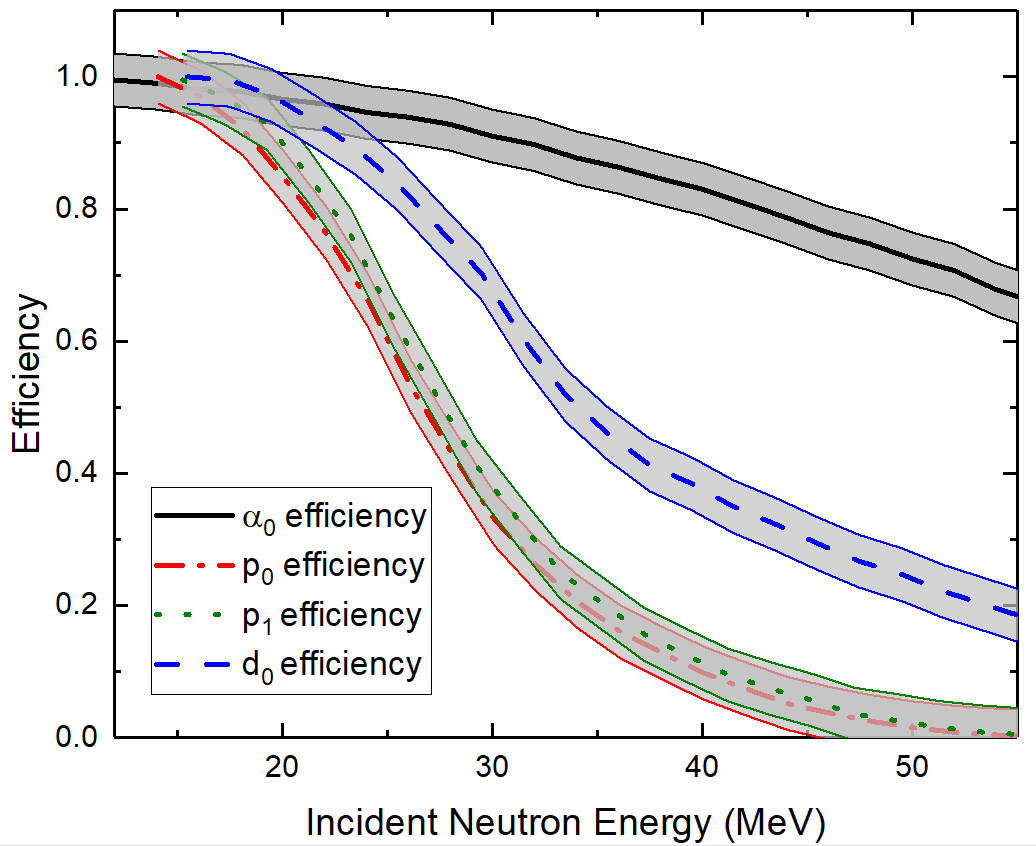}
\caption{\label{fig:Efficiency} Simulated detection efficiency as a function of incident neutron energy for each of the reactions studied. The different curves correspond to the detection efficiency for the products of each of the reactions, $^{12}$C(n,$\alpha_0$)$^9$Be, $^{12}$C(n,p$_0$)$^{12}$B, $^{12}$C(n,p$_1$)$^{12}$B, and $^{12}$C(n,d$_0$)$^{11}$B, respectively. The efficiency was used to correct each cross section data point. The uncertainties in the efficiencies are included in the error bands.}
\end{figure}

Due to the small thickness of the diamond detectors (0.5 mm), there exists a significant possibility of incomplete energy deposition by the light ions created in the reaction channels of interest. To account for this, simulations were run using the MENATE$\_$R package within GEANT4. MENATE$\_$R assumes an isotropic distribution in the center-of-mass frame. Each reaction channel was simulated separately, and the probability of events to be in the full energy peak as a function of incident neutron energy was determined. The results of these simulations are shown in Fig. \ref{fig:Efficiency}. These efficiencies were used to correct the counts for each reaction channel and in each energy bin. 

\begin{table*}[htb]
\caption{\label{tab:Uncertainty2}
Example of uncertainties in the cross sections. For the (n,$\alpha$) and (n,p$_0$) reactions, the uncertainty is included for the point with the least uncertainty and a point with larger uncertainty. The total uncertainty is the quadrature sum of the individual uncertainties, including statistical, charged-particle efficiency correction, fit/integration, previous data for normalization, and flux shape. Since they are common to each of the cross sections, the uncertainties in incident neutron energy reconstruction (0.49\%), charged-particle detection efficiency correction (4\%), and of previous measurements for normalization (4.9\%) are not included in the table. The individual uncertainties for the (n,d$_0$) and (n,p$_1$) reactions are similar to those of the (n,p$_0$). The statistical uncertainty listed here is the total statistical uncertainty, and includes contributions from the peak ($N$ in eq. 3), from the number of (n,$\alpha$) counts at 14.1 MeV, ($N_{n,\alpha}$ in eq. 3), and from the background counts for (n,p$_0$).}
\begin{ruledtabular}
\begin{tabular}{|c|c|c|c|c|c|c|}
\textbf{Reaction channel} & \textbf{E$_n$ (MeV)}  & \textbf{Statistical}& \textbf{Fit} & \textbf{Flux ($\Phi_\text{FC}$) } & \textbf{Total} & \textbf{Total uncertainty (mb)}\\ \hline
$^{12}$C(n,$\alpha_0$)$^9$Be  & 12.53 & 3.9\% & -  & 5\% & 8.9\% & 8.4\\ 
$^{12}$C(n,$\alpha_0$)$^9$Be  & 50.04 & 40.9\% & - & 23.2\% & 47.5\% & 0.1\\
$^{12}$C(n,p$_0$)$^{12}$B  & 17.36 & 7.9\% & 3.6\%  & 5.3\% & 11.1\% & 2.3\\ 
$^{12}$C(n,p$_0$)$^{12}$B  & 26.79 & 25.8\% & 7.8\% & 11.1\% & 27.2\% & 1.0\\

\end{tabular}
\end{ruledtabular}
\end{table*} 

\subsection{Uncertainty}
Various uncertainties contribute to the total uncertainty for these measurements. First, there is approximately 4.9\% uncertainty from the weighted average $^{12}$C(n,$\alpha_0$)$^9$Be cross section results from previous data that was used to normalize this work. The uncertainty in the neutron flux was derived from the differences between the fission chamber data and an MCNP simulation, starting at 5\% uncertainty from 12-17 MeV and gradually increasing to 20\% at higher energies. An uncertainty of 4\% arises from the charged particle detection efficiency correction, due to a lack of angular distribution data for the channels of interest. Statistical uncertainties were present as well and generally increase with increasing incident neutron energy. Starting at about 2.5\% for the lowest-energy (n,$\alpha$) data, these statistical uncertainties increase to over 20\% at higher energies. For channels besides $^{12}$C(n,$\alpha_0$)$^9$Be, fit  uncertainties are present due to the need to fit the peak and the background to extract the counts. An uncertainty of 0.49\% is included for the E$_n$ reconstruction uncertainty; this is the maximum incident energy uncertainty considering the time resolution and flight path length, and all other uncertainties on the E$_n$ reconstruction are lower. This is also used for the horizontal error bars for the cross section plots. These uncertainties were summed in quadrature to find the total uncertainty for each measured cross section. A summary of the uncertainties is included in Table \ref{tab:Uncertainty2} for a few sample data points. The full uncertainty quantification, as well as raw data and important correction factors, is included in the supplemental material.

\section{Results}
Cross sections were determined for four neutron-induced reactions on $^{12}$C: $^{12}$C(n,$\alpha_0$)$^9$Be, $^{12}$C(n,p$_0$)$^{12}$B, $^{12}$C(n,p$_1$)$^{12}$B, and $^{12}$C(n,d$_0$)$^{11}$B. The results for each of these are discussed separately in this section.

\begin{figure*}
\includegraphics[width=0.95\textwidth]{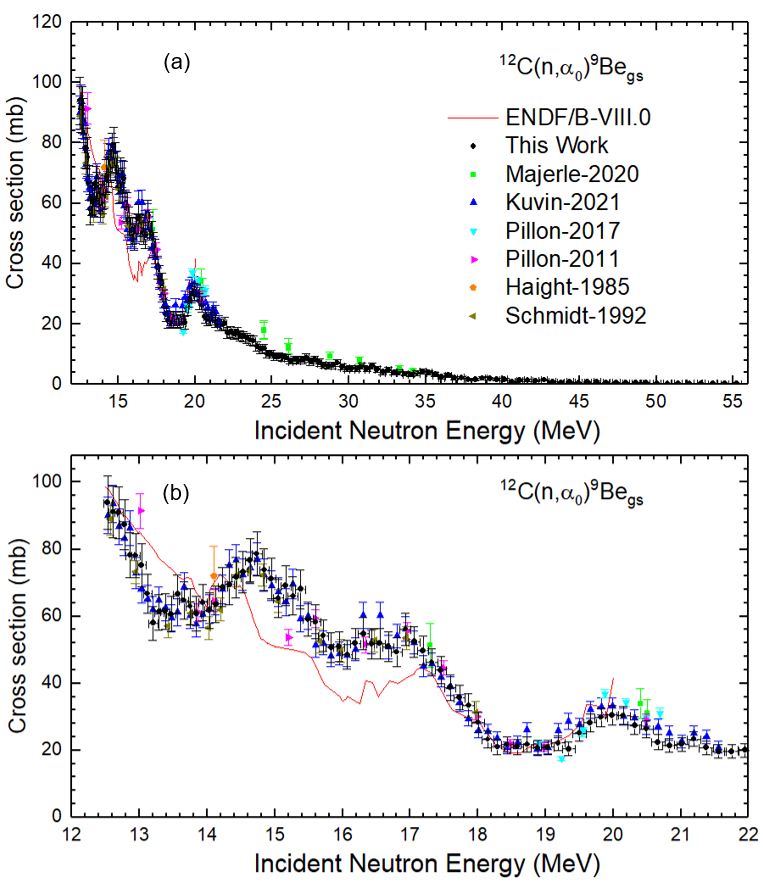}
\caption{\label{fig:n_alpha} Cross sections for the $^{12}$C(n,$\alpha_0$)$^{9}$Be reaction. The counts were normalized to previous $^{12}$C(n,$\alpha_0$)$^9$Be experimental data  at 14.1 MeV. This works shows excellent agreement with recent data \cite{Kuvin2021,MAJERLE2020163014,Pillon2011,Pillon2017, Haight1984,Schmidt1992}, but demonstrates only partial agreement with the ENDF evaluation. (a) $^{12}$C(n,$\alpha_0$)$^{9}$Be cross sections for the full range of our data, from 12.5 to 55 MeV. (b) Zoomed in view of (a), showing more clearly the previous data from 12.5-22 MeV, including the ENDF/B-VIII.0 evaluation.}
\end{figure*}

\subsection{\boldmath  $^{12}$C(n,$\alpha_0$)$^9$Be}
Cross sections for the $^{12}$C(n,$\alpha_0$)$^9$Be reaction were determined up to incident neutron energies of 55 MeV, higher than previous measurements which ended at 34 MeV \cite{MAJERLE2020163014}. The full extent of our dataset from 12.5 to 55 MeV is included in Fig. \ref{fig:n_alpha}a. A zoomed-in view from 12.5-22 MeV is shown in Fig. \ref{fig:n_alpha}b, where the bulk of the previous data exists. The cross sections and associated uncertainty are included in Table \ref{tab:table_alpha0_3col}. Excellent agreement is shown with recent data, particularly the four most recent studies, including the works of Pillon et al. (2011 and 2017) \cite{Pillon2011,Pillon2017}, Majerle et al. (2020) \cite{Pillon2017}, and Kuvin et al. (2021) \cite{Kuvin2021}, which all used diamond detectors and took advantage of their unique active target capabilities.  The ENDF/B-VIII.0 evaluation for this channel is shown for comparison. This work, in addition to the other recent experiments, shows the need to update the ENDF evaluation for this reaction. While the data from both diamond detectors were used for this channel, we used only the detector with better resolution (upstream) for the other channels, as it minimized the total uncertainty in the cross sections.

\subsection{\boldmath  $^{12}$C(n,p$_0$)$^{12}$B}
\begin{figure}
\includegraphics[width=\columnwidth]{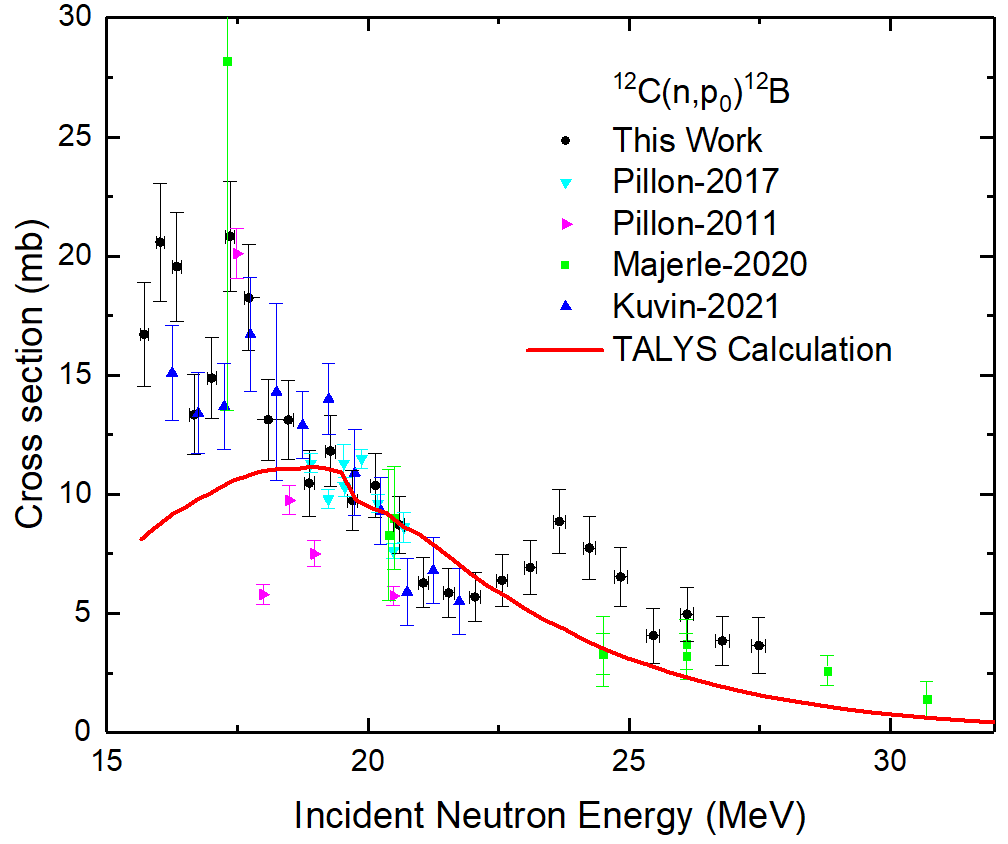}
\caption{\label{fig:n_p0} Cross sections for the $^{12}$C(n,p$_0$)$^{12}$B reaction, normalized to previously measured $^{12}$C(n,$\alpha_0$)$^9$Be experimental data at 14.1 MeV. Current and previous experimental data are compared to a TALYS calculation with default parameters.}
\end{figure}

Cross sections were also determined for the $^{12}$C(n,p$_0$)$^{12}$B reaction from slightly above the reaction threshold (13.646 MeV) at 15.7 MeV up to 27.5 MeV, shown in Table \ref{tab:table_p0_3col}. Figure \ref{fig:n_p0} shows a comparison with the other datasets that have published partial cross sections for producing a proton and $^{12}$B in the ground state. The same data labels are used for this and the other cross-section plots for consistency. Overall, our data agree reasonably well with previous data. In particular, the new data added above 22 MeV fill in the gaps left by previous studies. There is no ENDF evaluation for this partial cross section as there is for the $^{12}$C(n,$\alpha_0$)$^9$Be reaction, and this partial cross section is not directly comparable to the evaluation for the $^{12}$C(n,p) inclusive cross section. If all of the other proton-producing channels were probed, these could be summed together to make a comparison with the evaluation. Instead, a TALYS \cite{TALYS} calculation was performed using default parameters, as was done in \cite{Kuvin2021}, to test how the scale of the data compares. However, while the calculation matches the general trend of the data, such statistical calculations are not expected to reproduce the fluctuations in the cross section due to resonant behavior. The detection of this channel at higher incident neutron energies is especially limited by the low detection efficiency for the protons from this reaction. 

\subsection{\boldmath  $^{12}$C(n,p$_1$)$^{12}$B}
\begin{figure}
\includegraphics[width=\columnwidth]{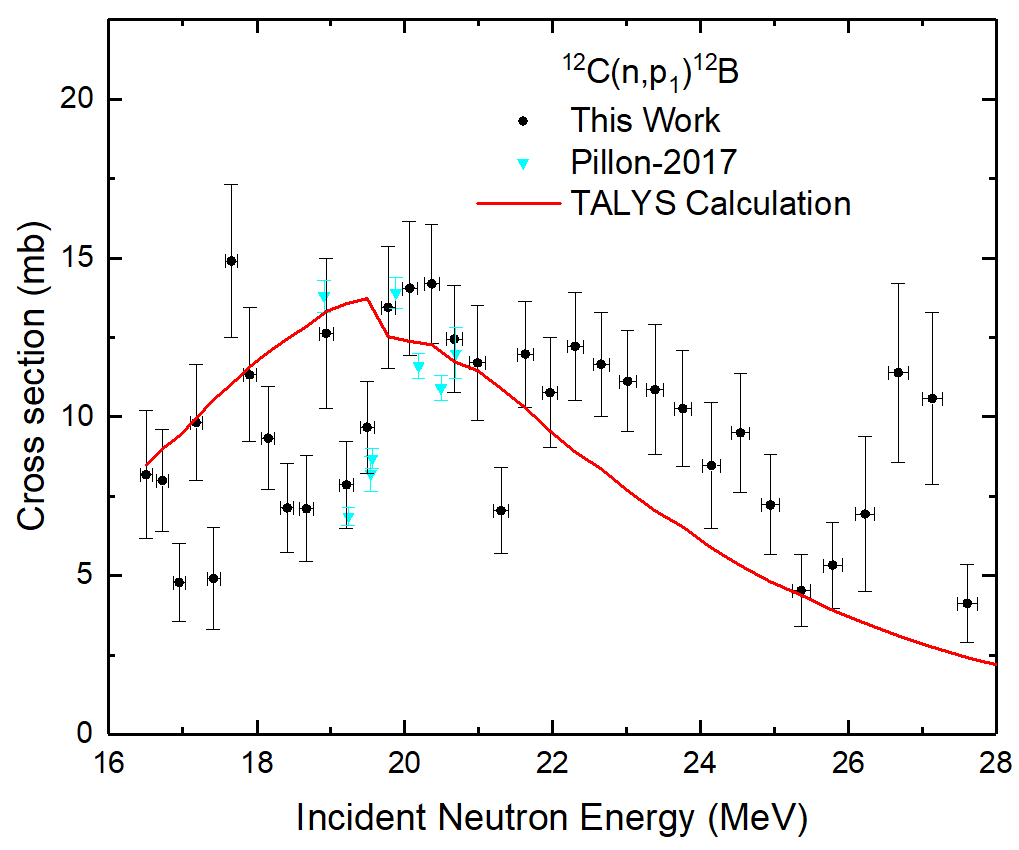}
\caption{\label{fig:n_p1} Cross sections for the $^{12}$C(n,p$_1$)$^{12}$B reaction, (populating $^{12}$B in the 2$^+$, 0.953 MeV first excited state) normalized to previously measured $^{12}$C(n,$\alpha_0$)$^9$Be experimental data at 14.1 MeV. Current and previous experimental data are compared to a TALYS calculation with default parameters.}
\end{figure}

In addition, cross sections were determined for the $^{12}$C(n,p$_1$)$^{12}$B reaction from 16.5 to 27.6 MeV, and are shown in Fig. \ref{fig:n_p1} and Table \ref{tab:table_p1_3col}. The isolation of this channel from the $^{12}$C(n,d$_0$)$^{11}$B channel requires excellent energy resolution, and as such there are fewer studies that could measure partial cross sections for producing a proton and $^{12}$B in the first excited state (2$^+$, E$_x$ = 0.953 MeV), only the work of Pillon et al. \cite{Pillon2017} and the present work. Another recent study, \cite{Kuvin2021}, published a combined (n,d$_0$+p$_1$) cross section, recognizing that with their resolution, the (n,p$_1$) could not be distinguished from the (n,d$_0$) due to their similar Q values (-13.54 and -13.73 MeV, respectively). Other studies that analyzed the (n,d$_0$) channel using a diamond detector will have (n,p$_1$) contamination unless they have sufficient resolution to isolate the contributions from the two peaks. Our data agree with the overall trend of Pillon's study, despite the somewhat limited statistics for this channel in the current work. However, the energy range in the current study extends significantly beyond that of previous studies. The scale of the TALYS calculation is in general agreement with the trend of the data, but again is not expected to account for any resonant behavior.

\subsection{\boldmath  $^{12}$C(n,d$_0$)$^{11}$B}
\begin{figure}
\includegraphics[width=\columnwidth]{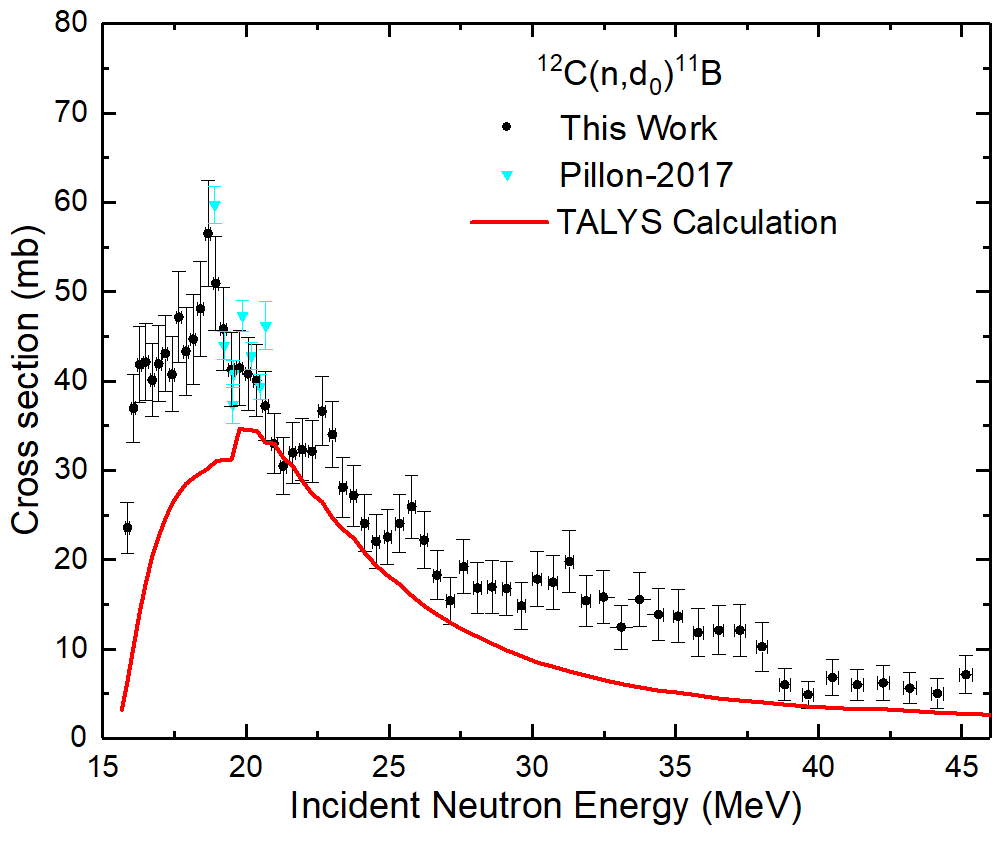}
\caption{\label{fig:n_d0} Cross sections for the $^{12}$C(n,d$_0$)$^{11}$B reaction, normalized to previously measured $^{12}$C(n,$\alpha_0$)$^9$Be experimental data at 14.1 MeV. Current and previous experimental data are compared to a TALYS calculation with default parameters.}
\end{figure}

The last reaction channel for which cross sections were obtained is $^{12}$C(n,d$_0$)$^{11}$B. These data are plotted in Fig. \ref{fig:n_d0}, and included in Table \ref{tab:table_d0_3col}. As mentioned previously, this channel's Q value matches that of the $^{12}$C(n,p$_1$)$^{12}$B channel closely. Due to other studies \cite{Pillon2011,MAJERLE2020163014,Kuvin2021}  having insufficient resolution to isolate the contributions from the two peaks, again only the Pillon dataset of 2017 \cite{Pillon2017} is directly comparable to the current study (i.e., it is a measurement of exclusively the production of a deuteron and $^{11}$B in the ground state). Although the (n,d$_0$) channel has a much larger cross section than the (n,p$_1$), the non-negligible contribution of the (n,p$_1$) channel ensures these other studies are still not directly comparable. 

The measured cross sections show reasonably good agreement with existing data, and new data are added below (down to 15.8 MeV) and above (up to 46 MeV). Again, a TALYS calculation of the cross section for this reaction channel using default parameters is included for comparison. The general shape of the TALYS calculation matches fairly well, but the scale does not agree with the data presented here, and the resonance structures are not reproduced.

\section{Conclusions}
Partial cross sections were determined for various neutron-induced reactions on $^{12}$C, including $^{12}$C(n,$\alpha_0$)$^9$Be, $^{12}$C(n,p$_0$)$^{12}$B, $^{12}$C(n,p$_1$)$^{12}$B, and $^{12}$C(n,d$_0$)$^{11}$B. These measurements significantly extend the energy ranges over which these cross sections have been measured. These newly measured cross sections will be used to improve the neutron modeling code MENATE\_R, and will eventually allow for better simulations of neutron interactions with carbon.  

This work also reinforces the need for updated ENDF evaluations for these reaction channels. The evaluation for the $^{12}$C(n,$\alpha_0$)$^9$Be channel needs to be updated to take into consideration the more recent data with diamond detectors. Evaluations of exclusive cross sections for  $^{12}$C(n,p$_0$)$^{12}$B, $^{12}$C(n,p$_1$)$^{12}$B, and $^{12}$C(n,d$_0$)$^{11}$B should be performed as well. These new evaluations would be beneficial for benchmarking future neutron experiments and improving simulations that depend on these data. 

\begin{acknowledgments}
 This work was supported by the U.S. Department of Energy through the Los Alamos National Laboratory, Contract No. 89233218CNA000001. This work would not have been possible without the LANSCE accelerator facility and the dedicated staff who operate it. This work was also supported by the National Science Foundation, grant numbers PHY 2012040, PHY 2310078, PHY 2311125, PHY 2311126, and PHY 2011265, and the U.S. Department of Energy, Office of Science, Office of Nuclear Physics, Award Number DE-SC0022037. 

\end{acknowledgments}

\begin{table*}[h]
\caption{\label{tab:table_alpha0_3col}
Partial cross sections and uncertainties for $^{12}$C(n,$\alpha_0$)$^{9}$Be, normalized to previously measured $^{12}$C(n,$\alpha_0$)$^9$Be experimental data at 14.1 MeV}
\begin{ruledtabular}
\begin{tabular}{c c p{0.0 in}| c c p{0.0 in} | c c }
 E$_n$ (MeV)& $\sigma$ (mb) &  & E$_n$ (MeV)& $\sigma$ (mb) & & E$_n$ (MeV) & $\sigma$ (mb) \\ 
 \hline
12.53  &  93.9  $\pm$  8.4  & &  18.29  &  21.0  $\pm$  2.4  & &  29.59  &  5.8  $\pm$  1.0  \\
12.61  &  91.1  $\pm$  8.1  & &  18.43  &  21.7  $\pm$  2.4  & &  29.90  &  5.1  $\pm$  0.9  \\
12.69  &  91.0  $\pm$  8.1  & &  18.58  &  21.0  $\pm$  2.3  & &  30.22  &  5.2  $\pm$  0.9  \\
12.78  &  87.4  $\pm$  7.8  & &  18.73  &  21.7  $\pm$  2.4  & &  30.54  &  5.2  $\pm$  0.9  \\
12.86  &  78.2  $\pm$  7.0  & &  18.88  &  20.7  $\pm$  2.2  & &  30.86  &  6.1  $\pm$  1.0  \\
12.94  &  78.1  $\pm$  7.0  & &  19.03  &  20.9  $\pm$  2.2  & &  31.20  &  5.1  $\pm$  0.9  \\
13.03  &  75.3  $\pm$  6.8  & &  19.19  &  22.1  $\pm$  2.3  & &  31.53  &  6.0  $\pm$  1.1  \\
13.12  &  66.7  $\pm$  6.1  & &  19.34  &  20.4  $\pm$  2.1  & &  31.88  &  5.0  $\pm$  0.9  \\
13.20  &  57.9  $\pm$  5.3  & &  19.50  &  25.1  $\pm$  2.5  & &  32.22  &  4.0  $\pm$  0.7  \\
13.29  &  61.2  $\pm$  5.6  & &  19.66  &  28.2  $\pm$  2.8  & &  32.58  &  4.7  $\pm$  0.9  \\
13.38  &  61.5  $\pm$  5.6  & &  19.82  &  29.8  $\pm$  2.9  & &  32.94  &  3.9  $\pm$  0.8  \\
13.47  &  60.6  $\pm$  5.5  & &  19.98  &  30.5  $\pm$  3.0  & &  33.30  &  3.9  $\pm$  0.8  \\
13.56  &  66.6  $\pm$  6.0  & &  20.15  &  30.3  $\pm$  3.0  & &  33.68  &  3.4  $\pm$  0.7  \\
13.66  &  64.7  $\pm$  5.9  & &  20.32  &  27.5  $\pm$  2.7  & &  34.06  &  3.3  $\pm$  0.7  \\
13.75  &  62.9  $\pm$  5.7  & &  20.49  &  26.3  $\pm$  2.6  & &  34.44  &  3.7  $\pm$  0.8  \\
13.84  &  60.7  $\pm$  5.5  & &  20.66  &  22.4  $\pm$  2.3  & &  34.83  &  4.1  $\pm$  0.8  \\
13.94  &  64.1  $\pm$  5.8  & &  20.83  &  21.4  $\pm$  2.2  & &  35.23  &  3.8  $\pm$  0.8  \\
14.03  &  61.8  $\pm$  5.6  & &  21.01  &  21.7  $\pm$  2.2  & &  35.64  &  3.3  $\pm$  0.7  \\
14.13  &  63.6  $\pm$  5.8  & &  21.19  &  23.3  $\pm$  2.3  & &  36.05  &  2.3  $\pm$  0.5  \\
14.23  &  68.7  $\pm$  6.2  & &  21.37  &  20.7  $\pm$  2.1  & &  36.47  &  2.9  $\pm$  0.6  \\
14.33  &  69.4  $\pm$  6.3  & &  21.56  &  19.6  $\pm$  2.0  & &  36.90  &  2.0  $\pm$  0.5  \\
14.43  &  71.6  $\pm$  6.5  & &  21.75  &  19.7  $\pm$  2.1  & &  37.34  &  2.1  $\pm$  0.5  \\
14.53  &  73.2  $\pm$  6.6  & &  21.95  &  20.1  $\pm$  2.1  & &  37.78  &  1.6  $\pm$  0.4  \\
14.63  &  76.7  $\pm$  6.9  & &  22.15  &  17.0  $\pm$  1.8  & &  38.23  &  1.5  $\pm$  0.4  \\
14.74  &  78.6  $\pm$  7.1  & &  22.35  &  17.2  $\pm$  1.8  & &  38.70  &  2.0  $\pm$  0.5  \\
14.84  &  73.8  $\pm$  6.7  & &  22.55  &  16.6  $\pm$  1.8  & &  39.16  &  1.9  $\pm$  0.5  \\
14.95  &  71.2  $\pm$  6.4  & &  22.75  &  17.0  $\pm$  1.8  & &  39.64  &  1.5  $\pm$  0.4  \\
15.05  &  65.3  $\pm$  5.9  & &  22.96  &  15.9  $\pm$  1.7  & &  40.13  &  1.6  $\pm$  0.4  \\
15.16  &  69.1  $\pm$  6.3  & &  23.17  &  15.8  $\pm$  1.8  & &  40.62  &  0.9  $\pm$  0.2  \\
15.27  &  66.0  $\pm$  6.0  & &  23.39  &  15.3  $\pm$  1.8  & &  41.13  &  1.3  $\pm$  0.3  \\
15.38  &  68.0  $\pm$  6.2  & &  23.60  &  14.6  $\pm$  1.8  & &  41.65  &  1.3  $\pm$  0.3  \\
15.49  &  59.1  $\pm$  5.4  & &  23.83  &  14.4  $\pm$  1.8  & &  42.17  &  1.2  $\pm$  0.3  \\
15.60  &  58.4  $\pm$  5.3  & &  24.05  &  12.5  $\pm$  1.6  & &  42.71  &  1.4  $\pm$  0.4  \\
15.72  &  54.1  $\pm$  5.0  & &  24.28  &  11.7  $\pm$  1.5  & &  43.25  &  0.7  $\pm$  0.2  \\
15.83  &  50.7  $\pm$  4.7  & &  24.51  &  11.8  $\pm$  1.5  & &  43.81  &  0.6  $\pm$  0.2  \\
15.95  &  50.9  $\pm$  4.7  & &  24.74  &  10.1  $\pm$  1.3  & &  44.38  &  0.8  $\pm$  0.2  \\
16.07  &  48.5  $\pm$  4.5  & &  24.98  &  9.9  $\pm$  1.3  & &  44.95  &  0.5  $\pm$  0.2  \\
16.19  &  51.8  $\pm$  4.8  & &  25.22  &  9.2  $\pm$  1.3  & &  45.55  &  0.7  $\pm$  0.2  \\
16.31  &  54.8  $\pm$  5.0  & &  25.46  &  9.2  $\pm$  1.3  & &  46.15  &  0.8  $\pm$  0.2  \\
16.43  &  51.8  $\pm$  4.8  & &  25.71  &  8.8  $\pm$  1.2  & &  46.76  &  0.5  $\pm$  0.2  \\
16.55  &  51.9  $\pm$  4.8  & &  25.96  &  8.7  $\pm$  1.2  & &  47.39  &  0.5  $\pm$  0.2  \\
16.68  &  50.8  $\pm$  4.7  & &  26.21  &  7.7  $\pm$  1.1  & &  48.03  &  0.4  $\pm$  0.1  \\
16.80  &  49.3  $\pm$  4.5  & &  26.47  &  8.2  $\pm$  1.2  & &  48.69  &  0.4  $\pm$  0.1  \\
16.93  &  56.1  $\pm$  5.1  & &  26.73  &  7.8  $\pm$  1.2  & &  49.35  &  0.4  $\pm$  0.1  \\
17.06  &  52.4  $\pm$  4.8  & &  27.00  &  8.1  $\pm$  1.3  & &  50.04  &  0.2  $\pm$  0.1  \\
17.19  &  49.7  $\pm$  4.6  & &  27.27  &  8.7  $\pm$  1.4  & &  50.73  &  0.3  $\pm$  0.1  \\
17.32  &  46.2  $\pm$  4.3  & &  27.54  &  7.5  $\pm$  1.2  & &  51.44  &  0.2  $\pm$  0.1  \\
17.45  &  43.9  $\pm$  4.3  & &  27.82  &  8.1  $\pm$  1.3  & &  52.17  &  0.2  $\pm$  0.1  \\
17.59  &  38.8  $\pm$  4.0  & &  28.11  &  7.2  $\pm$  1.2  & &  52.92  &  0.2  $\pm$  0.1  \\
17.73  &  35.7  $\pm$  3.8  & &  28.39  &  6.3  $\pm$  1.1  & &  53.67  &  0.1  $\pm$  0.1  \\
17.86  &  33.3  $\pm$  3.7  & &  28.69  &  6.1  $\pm$  1.1  & &  54.45  &  0.1  $\pm$  0.1  \\
18.00  &  28.3  $\pm$  3.2  & &  28.98  &  6.2  $\pm$  1.1  & &  55.25  &  0.1  $\pm$  0.1  \\
18.15  &  23.3  $\pm$  2.6  & &  29.28  &  7.0  $\pm$  1.2  & &  \\

\end{tabular}
\end{ruledtabular}
\end{table*}

\begin{table*}[h]
\caption{\label{tab:table_p0_3col}
Partial cross sections and uncertainties for $^{12}$C(n,p$_0$)$^{12}$B, normalized to previously measured $^{12}$C(n,$\alpha_0$)$^9$Be experimental data at 14.1 MeV}
\begin{ruledtabular}
\begin{tabular}{c c p{0.0 in}| c c p{0.0 in} | c c }
 E$_n$ (MeV) & $\sigma$ (mb) &  & E$_n$ (MeV) & $\sigma$ (mb) & & E$_n$ (MeV) & $\sigma$ (mb) \\ 
 \hline
15.71  &  16.7  $\pm$  2.2  & &  18.87  &  10.5  $\pm$  1.4  & &  23.11  &  6.9  $\pm$  1.1  \\
16.02  &  20.6  $\pm$  2.5  & &  19.28  &  11.8  $\pm$  1.5  & &  23.66  &  8.9  $\pm$  1.4  \\
16.34  &  19.6  $\pm$  2.3  & &  19.71  &  9.7  $\pm$  1.3  & &  24.24  &  7.7  $\pm$  1.3  \\
16.67  &  13.3  $\pm$  1.7  & &  20.14  &  10.4  $\pm$  1.4  & &  24.84  &  6.5  $\pm$  1.3  \\
17.01  &  14.9  $\pm$  1.8  & &  20.6  &  8.7  $\pm$  1.2  & &  25.47  &  4.1  $\pm$  1.2  \\
17.36  &  20.8  $\pm$  2.4  & &  21.06  &  6.3  $\pm$  1.1  & &  26.11  &  5  $\pm$  1.1  \\
17.72  &  18.2  $\pm$  2.3  & &  21.55  &  5.9  $\pm$  1  & &  26.79  &  3.8  $\pm$  1.1  \\
18.09  &  13.1  $\pm$  1.7  & &  22.05  &  5.7  $\pm$  1  & &  27.49  &  3.7  $\pm$  1.2  \\
18.48  &  13.1  $\pm$  1.7  & &  22.57  &  6.4  $\pm$  1.1  & &  \\
\end{tabular}
\end{ruledtabular}
\end{table*}

\begin{table*}[h]
\caption{\label{tab:table_p1_3col}
Partial cross sections and uncertainties for $^{12}$C(n,p$_1$)$^{12}$B, normalized to previously measured $^{12}$C(n,$\alpha_0$)$^9$Be experimental data at 14.1 MeV}
\begin{ruledtabular}
\begin{tabular}{c c p{0.0 in}| c c p{0.0 in} | c c }
 E$_n$ (MeV) & $\sigma$ (mb) &  & E$_n$ (MeV) & $\sigma$ (mb) & & E$_n$ (MeV) & $\sigma$ (mb) \\ 
 \hline
16.5  &  8.2  $\pm$  2  & &  19.49  &  9.7  $\pm$  1.4  & &  23.38  &  10.9  $\pm$  2.1  \\
16.73  &  8  $\pm$  1.6  & &  19.78  &  13.4  $\pm$  1.9  & &  23.76  &  10.3  $\pm$  1.8  \\
16.95  &  4.8  $\pm$  1.2  & &  20.07  &  14  $\pm$  2.1  & &  24.14  &  8.5  $\pm$  2  \\
17.18  &  9.8  $\pm$  1.8  & &  20.37  &  14.2  $\pm$  1.9  & &  24.54  &  9.5  $\pm$  1.9  \\
17.42  &  4.9  $\pm$  1.6  & &  20.67  &  12.4  $\pm$  1.7  & &  24.94  &  7.2  $\pm$  1.6  \\
17.66  &  14.9  $\pm$  2.4  & &  20.98  &  11.7  $\pm$  1.8  & &  25.36  &  4.5  $\pm$  1.1  \\
17.9  &  11.3  $\pm$  2.1  & &  21.3  &  7.1  $\pm$  1.4  & &  25.79  &  5.3  $\pm$  1.3  \\
18.15  &  9.3  $\pm$  1.6  & &  21.63  &  12  $\pm$  1.7  & &  26.22  &  6.9  $\pm$  2.4  \\
18.41  &  7.1  $\pm$  1.4  & &  21.96  &  10.8  $\pm$  1.7  & &  26.67  &  11.4  $\pm$  2.8  \\
18.67  &  7.1  $\pm$  1.7  & &  22.31  &  12.2  $\pm$  1.7  & &  27.13  &  10.6  $\pm$  2.7  \\
18.94  &  12.6  $\pm$  2.4  & &  22.66  &  11.7  $\pm$  1.6  & &  27.6  &  4.1  $\pm$  1.2  \\
19.21  &  7.9  $\pm$  1.4  & &  23.01  &  11.1  $\pm$  1.6  & &  \\

\end{tabular}
\end{ruledtabular}
\end{table*}

\begin{table*}[h]
\caption{\label{tab:table_d0_3col}
Partial cross sections and uncertainties for $^{12}$C(n,d$_0$)$^{11}$B, normalized to previously measured $^{12}$C(n,$\alpha_0$)$^9$Be experimental data at 14.1 MeV}
\begin{ruledtabular}
\begin{tabular}{c c p{0.0 in}| c c p{0.0 in} | c c }
 E$_n$ (MeV) & $\sigma$ (mb) &  & E$_n$ (MeV) & $\sigma$ (mb) & & E$_n$ (MeV) & $\sigma$ (mb) \\ 
 \hline
15.87  &  23.6  $\pm$  2.8  & &  21.3  &  30.5  $\pm$  3.2  & &  30.16  &  17.9  $\pm$  3.1  \\
16.07  &  37  $\pm$  3.8  & &  21.63  &  32  $\pm$  3.4  & &  30.72  &  17.5  $\pm$  3.1  \\
16.29  &  41.9  $\pm$  4.3  & &  21.96  &  32.3  $\pm$  3.5  & &  31.29  &  19.8  $\pm$  3.4  \\
16.5  &  42.1  $\pm$  4.3  & &  22.31  &  32.1  $\pm$  3.4  & &  31.88  &  15.4  $\pm$  2.8  \\
16.73  &  40.1  $\pm$  4.1  & &  22.66  &  36.6  $\pm$  3.8  & &  32.48  &  15.8  $\pm$  3  \\
16.95  &  41.9  $\pm$  4.2  & &  23.01  &  34  $\pm$  3.7  & &  33.1  &  12.5  $\pm$  2.5  \\
17.18  &  43.1  $\pm$  4.3  & &  23.38  &  28.1  $\pm$  3.3  & &  33.75  &  15.5  $\pm$  3.1  \\
17.42  &  40.8  $\pm$  4.2  & &  23.76  &  27.2  $\pm$  3.4  & &  34.41  &  13.9  $\pm$  3  \\
17.66  &  47.2  $\pm$  5.1  & &  24.14  &  24.1  $\pm$  3.2  & &  35.08  &  13.7  $\pm$  3  \\
17.9  &  43.4  $\pm$  5  & &  24.54  &  22.1  $\pm$  3  & &  35.79  &  11.9  $\pm$  2.7  \\
18.15  &  44.7  $\pm$  5.1  & &  24.94  &  22.5  $\pm$  3.1  & &  36.51  &  12.1  $\pm$  2.7  \\
18.41  &  48.1  $\pm$  5.3  & &  25.36  &  24  $\pm$  3.2  & &  37.25  &  12.1  $\pm$  2.9  \\
18.67  &  56.5  $\pm$  6  & &  25.79  &  26  $\pm$  3.5  & &  38.02  &  10.3  $\pm$  2.7  \\
18.94  &  50.9  $\pm$  5.3  & &  26.22  &  22.2  $\pm$  3.2  & &  38.81  &  6  $\pm$  1.8  \\
19.21  &  45.8  $\pm$  4.6  & &  26.67  &  18.3  $\pm$  2.7  & &  39.63  &  4.9  $\pm$  1.5  \\
19.49  &  41.3  $\pm$  4.1  & &  27.13  &  15.4  $\pm$  2.6  & &  40.47  &  6.8  $\pm$  2  \\
19.78  &  41.5  $\pm$  4.2  & &  27.6  &  19.2  $\pm$  3  & &  41.35  &  6  $\pm$  1.7  \\
20.07  &  40.8  $\pm$  4.1  & &  28.09  &  16.9  $\pm$  2.9  & &  42.25  &  6.2  $\pm$  2  \\
20.37  &  40.1  $\pm$  4  & &  28.59  &  17  $\pm$  3  & &  43.18  &  5.6  $\pm$  1.8  \\
20.67  &  37.2  $\pm$  3.8  & &  29.1  &  16.8  $\pm$  3  & &  44.14  &  5.1  $\pm$  1.7  \\
20.98  &  33  $\pm$  3.4  & &  29.62  &  14.9  $\pm$  2.6  & &  45.14  &  7.2  $\pm$  2.1  \\

\end{tabular}
\end{ruledtabular}
\end{table*}

\bibliography{apssamp}

\end{document}


\maketitle

\section{Data tables for each reaction channel}
\LTcapwidth=0.9\textwidth

\begin{longtable}[h]{|c|c|c|c|c|c|c|c|c|c|c|c|c|c|c|}
 \caption{$^{12}$C(n,$\alpha_0$)$^9$Be supplemental table. There was no background for this channel, so there is no statistical uncertainty due to the background. Additionally, because this channel had no background, the peaks were integrated instead of fit. Therefore, there are no fit uncertainties. The table contains the incident neutron energy ($E_n$), counts from $^{12}$C(n,$\alpha_0$)$^9$Be (Counts: Peak), the charged particle detection efficiency correction for this channel (Corrections: Efficiency), the simulated neutron transmission correction for the flux distribution (Corrections: Transmission), the cross section for this channel ($\sigma$ (mb)), the statistical uncertainties, broken into separate contributions from the counts from $^{12}$C(n,$\alpha_0$)$^9$Be at each energy (Statistical Uncertainty: $\sigma_{Peak}$) and from the (n,$\alpha$) counts at 14.1 MeV used for normalization (Statistical uncertainties: $\sigma_{N_{n,\alpha}}$), the total statistical uncertainties  combining all statistical uncertainties (Statistical Uncertainties: $\sigma_{stat. total}$), and systematic uncertainties, which include the charged particle detection efficiency correction uncertainty (Systematic Uncertainties: $\sigma_{Eff.}$), the relative flux distribution uncertainty (Systematic Uncertainties:  $\sigma_{Flux}$), the fit/integration uncertainties (Systematic Uncertainties: $\sigma_{Fit}$), the incident energy reconstruction uncertainty (systematic uncertainties: $\sigma_{NKE}$), and the normalization uncertainty based on the weighted average uncertainty of previous data (Systematic Uncertainties: $\sigma_{Norm.}$). The total uncertainty, which includes both the statistical and systematic uncertainties, is included as well, both as a percentage and in terms of millibarns. \label{n,alpha}}\\
 \hline
$E_n$  & Counts & \multicolumn{2}{c|}{Corrections} & Cross section & \multicolumn{3}{c|}{Statistical Uncertainties} &  \multicolumn{5}{c|}{Systematic Uncertainties} & Total Unc. & Error bar\\
\hline    
(MeV) & Peak & Efficiency & Transmission & $\sigma$ (mb) & $\sigma_{Peak}$ &   $\sigma_{N_{n,\alpha}}$ & $\sigma_{stat. total}$ & $\sigma_{Eff.}$ & $\sigma_{Flux}$ & $\sigma_{Fit}$ & $\sigma_{NKE}$ & $\sigma_{Norm.}$ & $\sigma_{Total}$ & Error (mb)\\
 \hline
 \endfirsthead
 \hline
$E_n$  & Counts & \multicolumn{2}{c|}{Corrections} & Cross section & \multicolumn{3}{c|}{Statistical Uncertainties} &  \multicolumn{5}{c|}{Systematic Uncertainties} & Total Unc. & Error bar\\
\hline
(MeV) & Counts & Efficiency & Transmission & $\sigma$ (mb) & $\sigma_{Peak}$ &   $\sigma_{N_{n,\alpha}}$ & $\sigma_{stat. total}$ & $\sigma_{Eff.}$ & $\sigma_{Flux}$ & $\sigma_{Fit}$ & $\sigma_{NKE}$ & $\sigma_{Norm.}$ & $\sigma_{Total}$ & Error (mb)\\ \hline
 \endhead

 \hline
 \endfoot
12.53 & 1627 & 99.3\% & 82.3\% & 93.9 & 2.48\% & 2.96\% & 3.86\% & 4\% & 5\% & - & 0.49\% & 4.87\% & 8.94\% & 8.4 \\
12.61 & 1612 & 99.5\% & 82.1\% & 91.1 & 2.49\% & 2.96\% & 3.87\% & 4\% & 5\% & - & 0.49\% & 4.87\% & 8.94\% & 8.1 \\
12.69 & 1641 & 99.2\% & 82.2\% & 91.0 & 2.47\% & 2.96\% & 3.86\% & 4\% & 5\% & - & 0.49\% & 4.87\% & 8.94\% & 8.1 \\
12.78 & 1506 & 99.2\% & 81.8\% & 87.4 & 2.58\% & 2.96\% & 3.93\% & 4\% & 5\% & - & 0.49\% & 4.87\% & 8.97\% & 7.8 \\
12.86 & 1330 & 99.1\% & 81.7\% & 78.2 & 2.74\% & 2.96\% & 4.04\% & 4\% & 5\% & - & 0.49\% & 4.87\% & 9.02\% & 7 \\
12.94 & 1390 & 99.1\% & 81.5\% & 78.1 & 2.68\% & 2.96\% & 4\% & 4\% & 5\% & - & 0.49\% & 4.87\% & 9\% & 7 \\
13.03 & 1333 & 99.3\% & 81.9\% & 75.3 & 2.74\% & 2.96\% & 4.04\% & 4\% & 5\% & - & 0.49\% & 4.87\% & 9.02\% & 6.8 \\
13.12 & 1132 & 99.3\% & 81.6\% & 66.7 & 2.97\% & 2.96\% & 4.2\% & 4\% & 5\% & - & 0.49\% & 4.87\% & 9.09\% & 6.1 \\
13.2 & 1002 & 99.2\% & 81.6\% & 57.9 & 3.16\% & 2.96\% & 4.33\% & 4\% & 5\% & - & 0.49\% & 4.87\% & 9.15\% & 5.3 \\
13.29 & 1084 & 99.3\% & 82\% & 61.2 & 3.04\% & 2.96\% & 4.24\% & 4\% & 5\% & - & 0.49\% & 4.87\% & 9.11\% & 5.6 \\
13.38 & 1076 & 99.2\% & 82.2\% & 61.5 & 3.05\% & 2.96\% & 4.25\% & 4\% & 5\% & - & 0.49\% & 4.87\% & 9.11\% & 5.6 \\
13.47 & 1068 & 99\% & 81.7\% & 60.6 & 3.06\% & 2.96\% & 4.26\% & 4\% & 5\% & - & 0.49\% & 4.87\% & 9.12\% & 5.5 \\
13.56 & 1198 & 99.1\% & 82.3\% & 66.6 & 2.89\% & 2.96\% & 4.14\% & 4\% & 5\% & - & 0.49\% & 4.87\% & 9.06\% & 6 \\
13.66 & 1119 & 99.1\% & 82.2\% & 64.7 & 2.99\% & 2.96\% & 4.21\% & 4\% & 5\% & - & 0.49\% & 4.87\% & 9.09\% & 5.9 \\
13.75 & 1075 & 99\% & 82.1\% & 62.9 & 3.05\% & 2.96\% & 4.25\% & 4\% & 5\% & - & 0.49\% & 4.87\% & 9.11\% & 5.7 \\
13.84 & 1056 & 99.2\% & 81.9\% & 60.7 & 3.08\% & 2.96\% & 4.27\% & 4\% & 5\% & - & 0.49\% & 4.87\% & 9.12\% & 5.5 \\
13.94 & 1119 & 99.2\% & 83.1\% & 64.1 & 2.99\% & 2.96\% & 4.21\% & 4\% & 5\% & - & 0.49\% & 4.87\% & 9.09\% & 5.8 \\
14.03 & 1088 & 99.1\% & 82.5\% & 61.8 & 3.03\% & 2.96\% & 4.24\% & 4\% & 5\% & - & 0.49\% & 4.87\% & 9.11\% & 5.6 \\
14.13 & 1139 & 98.8\% & 82.6\% & 63.6 & 2.96\% & 2.96\% & 4.19\% & 4\% & 5\% & - & 0.49\% & 4.87\% & 9.09\% & 5.8 \\
14.23 & 1224 & 99\% & 82.8\% & 68.7 & 2.86\% & 2.96\% & 4.12\% & 4\% & 5\% & - & 0.49\% & 4.87\% & 9.05\% & 6.2 \\
14.33 & 1230 & 99\% & 83\% & 69.4 & 2.85\% & 2.96\% & 4.11\% & 4\% & 5\% & - & 0.49\% & 4.87\% & 9.05\% & 6.3 \\
14.43 & 1247 & 99\% & 82.6\% & 71.6 & 2.83\% & 2.96\% & 4.1\% & 4\% & 5\% & - & 0.49\% & 4.87\% & 9.04\% & 6.5 \\
14.53 & 1292 & 98.8\% & 82.7\% & 73.2 & 2.78\% & 2.96\% & 4.06\% & 4\% & 5\% & - & 0.49\% & 4.87\% & 9.03\% & 6.6 \\
14.63 & 1370 & 98.9\% & 82.5\% & 76.7 & 2.7\% & 2.96\% & 4.01\% & 4\% & 5\% & - & 0.49\% & 4.87\% & 9\% & 6.9 \\
14.74 & 1379 & 98.9\% & 81.6\% & 78.6 & 2.69\% & 2.96\% & 4\% & 4\% & 5\% & - & 0.49\% & 4.87\% & 9\% & 7.1 \\
14.84 & 1289 & 98.7\% & 81.9\% & 73.8 & 2.79\% & 2.96\% & 4.07\% & 4\% & 5\% & - & 0.49\% & 4.87\% & 9.03\% & 6.7 \\
14.95 & 1228 & 98.5\% & 81.6\% & 71.2 & 2.85\% & 2.96\% & 4.11\% & 4\% & 5\% & - & 0.49\% & 4.87\% & 9.05\% & 6.4 \\
15.05 & 1138 & 98.4\% & 81.4\% & 65.3 & 2.96\% & 2.96\% & 4.19\% & 4\% & 5\% & - & 0.49\% & 4.87\% & 9.09\% & 5.9 \\
15.16 & 1207 & 98.5\% & 80.7\% & 69.1 & 2.88\% & 2.96\% & 4.13\% & 4\% & 5\% & - & 0.49\% & 4.87\% & 9.06\% & 6.3 \\
15.27 & 1153 & 98.5\% & 81.5\% & 66.0 & 2.95\% & 2.96\% & 4.18\% & 4\% & 5\% & - & 0.49\% & 4.87\% & 9.08\% & 6 \\
15.38 & 1172 & 98.7\% & 80.5\% & 68.0 & 2.92\% & 2.96\% & 4.16\% & 4\% & 5\% & - & 0.49\% & 4.87\% & 9.07\% & 6.2 \\
15.49 & 1026 & 98.7\% & 80.5\% & 59.1 & 3.12\% & 2.96\% & 4.3\% & 4\% & 5\% & - & 0.49\% & 4.87\% & 9.14\% & 5.4 \\
15.6 & 1018 & 98.4\% & 80.6\% & 58.4 & 3.13\% & 2.96\% & 4.31\% & 4\% & 5\% & - & 0.49\% & 4.87\% & 9.14\% & 5.3 \\
15.72 & 948 & 98.5\% & 80.7\% & 54.1 & 3.25\% & 2.96\% & 4.4\% & 4\% & 5\% & - & 0.49\% & 4.87\% & 9.18\% & 5 \\
15.83 & 890 & 98.2\% & 80.6\% & 50.7 & 3.35\% & 2.96\% & 4.47\% & 4\% & 5\% & - & 0.49\% & 4.87\% & 9.22\% & 4.7 \\
15.95 & 893 & 98.2\% & 80.1\% & 50.9 & 3.35\% & 2.96\% & 4.47\% & 4\% & 5\% & - & 0.49\% & 4.87\% & 9.22\% & 4.7 \\
16.07 & 851 & 98.3\% & 80.3\% & 48.5 & 3.43\% & 2.96\% & 4.53\% & 4\% & 5\% & - & 0.49\% & 4.87\% & 9.25\% & 4.5 \\
16.19 & 901 & 98\% & 80.4\% & 51.8 & 3.33\% & 2.96\% & 4.46\% & 4\% & 5\% & - & 0.49\% & 4.87\% & 9.21\% & 4.8 \\
16.31 & 960 & 98.2\% & 81.1\% & 54.8 & 3.23\% & 2.96\% & 4.38\% & 4\% & 5\% & - & 0.49\% & 4.87\% & 9.18\% & 5 \\
16.43 & 918 & 98\% & 81.5\% & 51.8 & 3.3\% & 2.96\% & 4.44\% & 4\% & 5\% & - & 0.49\% & 4.87\% & 9.2\% & 4.8 \\
16.55 & 931 & 98.2\% & 81.5\% & 51.9 & 3.28\% & 2.96\% & 4.42\% & 4\% & 5\% & - & 0.49\% & 4.87\% & 9.19\% & 4.8 \\
16.68 & 917 & 98\% & 81.7\% & 50.8 & 3.3\% & 2.96\% & 4.44\% & 4\% & 5\% & - & 0.49\% & 4.87\% & 9.2\% & 4.7 \\
16.8 & 898 & 98\% & 81.6\% & 49.3 & 3.34\% & 2.96\% & 4.46\% & 4\% & 5\% & - & 0.49\% & 4.87\% & 9.21\% & 4.5 \\
16.93 & 1022 & 97.9\% & 81.7\% & 56.1 & 3.13\% & 2.96\% & 4.31\% & 4\% & 5\% & - & 0.49\% & 4.87\% & 9.14\% & 5.1 \\
17.06 & 949 & 98\% & 82.1\% & 52.4 & 3.25\% & 2.96\% & 4.4\% & 4\% & 5\% & - & 0.49\% & 4.87\% & 9.18\% & 4.8 \\
17.19 & 891 & 98\% & 82.2\% & 49.7 & 3.35\% & 2.96\% & 4.47\% & 4\% & 5\% & - & 0.49\% & 4.87\% & 9.22\% & 4.6 \\
17.32 & 839 & 97.9\% & 82.2\% & 46.2 & 3.45\% & 2.96\% & 4.55\% & 4\% & 5.11\% & - & 0.49\% & 4.87\% & 9.32\% & 4.3 \\
17.45 & 812 & 97.6\% & 82.5\% & 43.9 & 3.51\% & 2.96\% & 4.59\% & 4\% & 5.75\% & - & 0.49\% & 4.87\% & 9.7\% & 4.3 \\
17.59 & 712 & 97.7\% & 81.9\% & 38.8 & 3.75\% & 2.96\% & 4.78\% & 4\% & 6.42\% & - & 0.49\% & 4.87\% & 10.2\% & 4 \\
17.73 & 652 & 97.8\% & 82.1\% & 35.7 & 3.92\% & 2.96\% & 4.91\% & 4\% & 6.98\% & - & 0.49\% & 4.87\% & 10.62\% & 3.8 \\
17.86 & 616 & 97.9\% & 81.8\% & 33.3 & 4.03\% & 2.96\% & 5\% & 4\% & 7.54\% & - & 0.49\% & 4.87\% & 11.04\% & 3.7 \\
18 & 531 & 97.8\% & 81.9\% & 28.3 & 4.34\% & 2.96\% & 5.25\% & 4\% & 7.59\% & - & 0.49\% & 4.87\% & 11.19\% & 3.2 \\
18.15 & 431 & 97.5\% & 81.7\% & 23.3 & 4.82\% & 2.96\% & 5.66\% & 4\% & 7.5\% & - & 0.49\% & 4.87\% & 11.32\% & 2.6 \\
18.29 & 384 & 97.1\% & 81.4\% & 21.0 & 5.1\% & 2.96\% & 5.9\% & 4\% & 7.41\% & - & 0.49\% & 4.87\% & 11.39\% & 2.4 \\
18.43 & 407 & 97.7\% & 81.7\% & 21.7 & 4.96\% & 2.96\% & 5.77\% & 4\% & 7.27\% & - & 0.49\% & 4.87\% & 11.24\% & 2.4 \\
18.58 & 395 & 97.3\% & 81.8\% & 21.0 & 5.03\% & 2.96\% & 5.84\% & 4\% & 7\% & - & 0.49\% & 4.87\% & 11.09\% & 2.3 \\
18.73 & 406 & 97.2\% & 81.7\% & 21.7 & 4.96\% & 2.96\% & 5.78\% & 4\% & 6.71\% & - & 0.49\% & 4.87\% & 10.88\% & 2.4 \\
18.88 & 387 & 97\% & 81.8\% & 20.7 & 5.08\% & 2.96\% & 5.88\% & 4\% & 6.39\% & - & 0.49\% & 4.87\% & 10.75\% & 2.2 \\
19.03 & 388 & 96.8\% & 81.6\% & 20.9 & 5.08\% & 2.96\% & 5.88\% & 4\% & 6.04\% & - & 0.49\% & 4.87\% & 10.54\% & 2.2 \\
19.19 & 407 & 97\% & 81.1\% & 22.1 & 4.96\% & 2.96\% & 5.77\% & 4\% & 5.68\% & - & 0.49\% & 4.87\% & 10.28\% & 2.3 \\
19.34 & 375 & 97.1\% & 81\% & 20.4 & 5.16\% & 2.96\% & 5.95\% & 4\% & 5.43\% & - & 0.49\% & 4.87\% & 10.24\% & 2.1 \\
19.5 & 462 & 96.6\% & 80.9\% & 25.1 & 4.65\% & 2.96\% & 5.52\% & 4\% & 5.46\% & - & 0.49\% & 4.87\% & 10.01\% & 2.5 \\
19.66 & 518 & 96.9\% & 80.6\% & 28.2 & 4.39\% & 2.96\% & 5.3\% & 4\% & 5.49\% & - & 0.49\% & 4.87\% & 9.91\% & 2.8 \\
19.82 & 557 & 97.1\% & 81\% & 29.8 & 4.24\% & 2.96\% & 5.17\% & 4\% & 5.46\% & - & 0.49\% & 4.87\% & 9.82\% & 2.9 \\
19.98 & 577 & 96.7\% & 81.1\% & 30.5 & 4.16\% & 2.96\% & 5.11\% & 4\% & 5.41\% & - & 0.49\% & 4.87\% & 9.76\% & 3 \\
20.15 & 575 & 96.3\% & 81.2\% & 30.3 & 4.17\% & 2.96\% & 5.12\% & 4\% & 5.35\% & - & 0.49\% & 4.87\% & 9.74\% & 3 \\
20.32 & 523 & 96.5\% & 81.5\% & 27.5 & 4.37\% & 2.96\% & 5.28\% & 4\% & 5.47\% & - & 0.49\% & 4.87\% & 9.89\% & 2.7 \\
20.49 & 502 & 96.4\% & 81.7\% & 26.3 & 4.46\% & 2.96\% & 5.36\% & 4\% & 5.65\% & - & 0.49\% & 4.87\% & 10.03\% & 2.6 \\
20.66 & 433 & 96.2\% & 81.9\% & 22.4 & 4.81\% & 2.96\% & 5.65\% & 4\% & 5.84\% & - & 0.49\% & 4.87\% & 10.29\% & 2.3 \\
20.83 & 418 & 96.3\% & 82\% & 21.4 & 4.89\% & 2.96\% & 5.72\% & 4\% & 5.6\% & - & 0.49\% & 4.87\% & 10.2\% & 2.2 \\
21.01 & 422 & 95.7\% & 81.9\% & 21.7 & 4.87\% & 2.96\% & 5.7\% & 4\% & 5.22\% & - & 0.49\% & 4.87\% & 9.98\% & 2.2 \\
21.19 & 460 & 95.8\% & 82.2\% & 23.3 & 4.66\% & 2.96\% & 5.52\% & 4\% & 5.19\% & - & 0.49\% & 4.87\% & 9.87\% & 2.3 \\
21.37 & 419 & 96.2\% & 82.1\% & 20.7 & 4.89\% & 2.96\% & 5.71\% & 4\% & 5.55\% & - & 0.49\% & 4.87\% & 10.17\% & 2.1 \\
21.56 & 404 & 96.2\% & 81.9\% & 19.6 & 4.98\% & 2.96\% & 5.79\% & 4\% & 5.92\% & - & 0.49\% & 4.87\% & 10.42\% & 2 \\
21.75 & 411 & 96\% & 82.6\% & 19.7 & 4.93\% & 2.96\% & 5.75\% & 4\% & 6.04\% & - & 0.49\% & 4.87\% & 10.47\% & 2.1 \\
21.95 & 419 & 95.8\% & 82.2\% & 20.1 & 4.89\% & 2.96\% & 5.71\% & 4\% & 6.1\% & - & 0.49\% & 4.87\% & 10.48\% & 2.1 \\
22.15 & 358 & 95.6\% & 82.6\% & 17.0 & 5.29\% & 2.96\% & 6.06\% & 4\% & 6.12\% & - & 0.49\% & 4.87\% & 10.69\% & 1.8 \\
22.35 & 361 & 95.9\% & 82\% & 17.2 & 5.26\% & 2.96\% & 6.04\% & 4\% & 6\% & - & 0.49\% & 4.87\% & 10.6\% & 1.8 \\
22.55 & 347 & 95.6\% & 82\% & 16.6 & 5.37\% & 2.96\% & 6.13\% & 4\% & 5.87\% & - & 0.49\% & 4.87\% & 10.58\% & 1.8 \\
22.75 & 361 & 95.7\% & 82.3\% & 17.0 & 5.26\% & 2.96\% & 6.04\% & 4\% & 6.09\% & - & 0.49\% & 4.87\% & 10.66\% & 1.8 \\
22.96 & 339 & 95.4\% & 81.9\% & 15.9 & 5.43\% & 2.96\% & 6.19\% & 4\% & 6.45\% & - & 0.49\% & 4.87\% & 10.95\% & 1.7 \\
23.17 & 340 & 94.9\% & 82.2\% & 15.8 & 5.42\% & 2.96\% & 6.18\% & 4\% & 6.81\% & - & 0.49\% & 4.87\% & 11.16\% & 1.8 \\
23.39 & 334 & 94.9\% & 82.1\% & 15.3 & 5.47\% & 2.96\% & 6.22\% & 4\% & 7.44\% & - & 0.49\% & 4.87\% & 11.58\% & 1.8 \\
23.6 & 324 & 94.7\% & 82\% & 14.6 & 5.56\% & 2.96\% & 6.3\% & 4\% & 8.12\% & - & 0.49\% & 4.87\% & 12.06\% & 1.8 \\
23.83 & 324 & 95.1\% & 81.9\% & 14.4 & 5.56\% & 2.96\% & 6.3\% & 4\% & 8.61\% & - & 0.49\% & 4.87\% & 12.4\% & 1.8 \\
24.05 & 284 & 94.5\% & 82.1\% & 12.5 & 5.93\% & 2.96\% & 6.63\% & 4\% & 8.95\% & - & 0.49\% & 4.87\% & 12.81\% & 1.6 \\
24.28 & 269 & 94.8\% & 81.9\% & 11.7 & 6.1\% & 2.96\% & 6.78\% & 4\% & 9.3\% & - & 0.49\% & 4.87\% & 13.13\% & 1.5 \\
24.51 & 271 & 94.3\% & 82\% & 11.8 & 6.07\% & 2.96\% & 6.76\% & 4\% & 9.27\% & - & 0.49\% & 4.87\% & 13.1\% & 1.5 \\
24.74 & 234 & 94.4\% & 82.7\% & 10.1 & 6.54\% & 2.96\% & 7.18\% & 4\% & 9.2\% & - & 0.49\% & 4.87\% & 13.27\% & 1.3 \\
24.98 & 229 & 94\% & 82.2\% & 9.9 & 6.61\% & 2.96\% & 7.24\% & 4\% & 9.34\% & - & 0.49\% & 4.87\% & 13.41\% & 1.3 \\
25.22 & 215 & 93.7\% & 82.7\% & 9.2 & 6.82\% & 2.96\% & 7.44\% & 4\% & 9.57\% & - & 0.49\% & 4.87\% & 13.67\% & 1.3 \\
25.46 & 215 & 93.8\% & 82.4\% & 9.2 & 6.82\% & 2.96\% & 7.44\% & 4\% & 9.78\% & - & 0.49\% & 4.87\% & 13.82\% & 1.3 \\
25.71 & 209 & 94\% & 82.6\% & 8.8 & 6.92\% & 2.96\% & 7.53\% & 4\% & 9.91\% & - & 0.49\% & 4.87\% & 13.96\% & 1.2 \\
25.96 & 207 & 93.9\% & 82.6\% & 8.7 & 6.95\% & 2.96\% & 7.56\% & 4\% & 10.05\% & - & 0.49\% & 4.87\% & 14.07\% & 1.2 \\
26.21 & 183 & 93.3\% & 82.6\% & 7.7 & 7.39\% & 2.96\% & 7.96\% & 4\% & 10.17\% & - & 0.49\% & 4.87\% & 14.38\% & 1.1 \\
26.47 & 195 & 93\% & 82.9\% & 8.2 & 7.16\% & 2.96\% & 7.75\% & 4\% & 10.29\% & - & 0.49\% & 4.87\% & 14.35\% & 1.2 \\
26.73 & 189 & 92.8\% & 83.2\% & 7.8 & 7.27\% & 2.96\% & 7.85\% & 4\% & 10.89\% & - & 0.49\% & 4.87\% & 14.84\% & 1.2 \\
27 & 200 & 92.5\% & 83.2\% & 8.1 & 7.07\% & 2.96\% & 7.67\% & 4\% & 11.98\% & - & 0.49\% & 4.87\% & 15.56\% & 1.3 \\
27.27 & 218 & 92.8\% & 83.4\% & 8.7 & 6.77\% & 2.96\% & 7.39\% & 4\% & 12.71\% & - & 0.49\% & 4.87\% & 16.01\% & 1.4 \\
27.54 & 186 & 92.4\% & 83.4\% & 7.5 & 7.33\% & 2.96\% & 7.91\% & 4\% & 12.35\% & - & 0.49\% & 4.87\% & 15.97\% & 1.2 \\
27.82 & 200 & 92.5\% & 83.3\% & 8.1 & 7.07\% & 2.96\% & 7.67\% & 4\% & 11.97\% & - & 0.49\% & 4.87\% & 15.56\% & 1.3 \\
28.11 & 179 & 92.6\% & 83.3\% & 7.2 & 7.47\% & 2.96\% & 8.04\% & 4\% & 12.58\% & - & 0.49\% & 4.87\% & 16.21\% & 1.2 \\
28.39 & 160 & 91.8\% & 83.3\% & 6.3 & 7.91\% & 2.96\% & 8.44\% & 4\% & 13.58\% & - & 0.49\% & 4.87\% & 17.2\% & 1.1 \\
28.69 & 157 & 91.4\% & 83.8\% & 6.1 & 7.98\% & 2.96\% & 8.51\% & 4\% & 13.86\% & - & 0.49\% & 4.87\% & 17.45\% & 1.1 \\
28.98 & 160 & 91.2\% & 83.6\% & 6.2 & 7.91\% & 2.96\% & 8.44\% & 4\% & 13.98\% & - & 0.49\% & 4.87\% & 17.51\% & 1.1 \\
29.28 & 183 & 91.3\% & 83.6\% & 7.0 & 7.39\% & 2.96\% & 7.96\% & 4\% & 14.04\% & - & 0.49\% & 4.87\% & 17.34\% & 1.2 \\
29.59 & 152 & 91.5\% & 83.8\% & 5.8 & 8.11\% & 2.96\% & 8.64\% & 4\% & 13.96\% & - & 0.49\% & 4.87\% & 17.59\% & 1 \\
29.9 & 135 & 91.1\% & 83.7\% & 5.1 & 8.61\% & 2.96\% & 9.1\% & 4\% & 13.88\% & - & 0.49\% & 4.87\% & 17.76\% & 0.9 \\
30.22 & 138 & 90.6\% & 83.7\% & 5.2 & 8.51\% & 2.96\% & 9.01\% & 4\% & 13.73\% & - & 0.49\% & 4.87\% & 17.6\% & 0.9 \\
30.54 & 138 & 90.6\% & 84.1\% & 5.2 & 8.51\% & 2.96\% & 9.01\% & 4\% & 13.57\% & - & 0.49\% & 4.87\% & 17.48\% & 0.9 \\
30.86 & 161 & 90\% & 83.8\% & 6.1 & 7.88\% & 2.96\% & 8.42\% & 4\% & 13.7\% & - & 0.49\% & 4.87\% & 17.28\% & 1 \\
31.2 & 139 & 90.1\% & 84.1\% & 5.1 & 8.48\% & 2.96\% & 8.98\% & 4\% & 13.91\% & - & 0.49\% & 4.87\% & 17.72\% & 0.9 \\
31.53 & 164 & 89.9\% & 84.2\% & 6.0 & 7.81\% & 2.96\% & 8.35\% & 4\% & 14.03\% & - & 0.49\% & 4.87\% & 17.51\% & 1.1 \\
31.88 & 136 & 89.5\% & 84.3\% & 5.0 & 8.57\% & 2.96\% & 9.07\% & 4\% & 14.12\% & - & 0.49\% & 4.87\% & 17.94\% & 0.9 \\
32.22 & 111 & 89.7\% & 84.3\% & 4.0 & 9.49\% & 2.96\% & 9.94\% & 4\% & 14.49\% & - & 0.49\% & 4.87\% & 18.68\% & 0.7 \\
32.58 & 132 & 89.1\% & 84.3\% & 4.7 & 8.7\% & 2.96\% & 9.19\% & 4\% & 15.02\% & - & 0.49\% & 4.87\% & 18.72\% & 0.9 \\
32.94 & 112 & 89\% & 84.7\% & 3.9 & 9.45\% & 2.96\% & 9.9\% & 4\% & 15.24\% & - & 0.49\% & 4.87\% & 19.25\% & 0.8 \\
33.3 & 113 & 88.7\% & 84.5\% & 3.9 & 9.41\% & 2.96\% & 9.86\% & 4\% & 15.3\% & - & 0.49\% & 4.87\% & 19.27\% & 0.8 \\
33.68 & 99 & 88.2\% & 84.7\% & 3.4 & 10.05\% & 2.96\% & 10.48\% & 4\% & 15.55\% & - & 0.49\% & 4.87\% & 19.79\% & 0.7 \\
34.06 & 97 & 87.7\% & 84.8\% & 3.3 & 10.15\% & 2.96\% & 10.58\% & 4\% & 16.04\% & - & 0.49\% & 4.87\% & 20.23\% & 0.7 \\
34.44 & 111 & 87.8\% & 85\% & 3.7 & 9.49\% & 2.96\% & 9.94\% & 4\% & 16.59\% & - & 0.49\% & 4.87\% & 20.35\% & 0.8 \\
34.83 & 125 & 87.8\% & 85\% & 4.1 & 8.94\% & 2.96\% & 9.42\% & 4\% & 17.24\% & - & 0.49\% & 4.87\% & 20.64\% & 0.8 \\
35.23 & 118 & 87.4\% & 84.9\% & 3.8 & 9.21\% & 2.96\% & 9.67\% & 4\% & 17.68\% & - & 0.49\% & 4.87\% & 21.13\% & 0.8 \\
35.64 & 100 & 86.6\% & 85.2\% & 3.3 & 10\% & 2.96\% & 10.43\% & 4\% & 17.72\% & - & 0.49\% & 4.87\% & 21.51\% & 0.7 \\
36.05 & 72 & 86.3\% & 85.2\% & 2.3 & 11.79\% & 2.96\% & 12.15\% & 4\% & 17.73\% & - & 0.49\% & 4.87\% & 22.41\% & 0.5 \\
36.47 & 90 & 85.4\% & 85.4\% & 2.9 & 10.54\% & 2.96\% & 10.95\% & 4\% & 17.97\% & - & 0.49\% & 4.87\% & 21.98\% & 0.6 \\
36.9 & 64 & 85.6\% & 85.2\% & 2.0 & 12.5\% & 2.96\% & 12.85\% & 4\% & 18.49\% & - & 0.49\% & 4.87\% & 23.39\% & 0.5 \\
37.34 & 69 & 85.3\% & 85.5\% & 2.1 & 12.04\% & 2.96\% & 12.4\% & 4\% & 19.32\% & - & 0.49\% & 4.87\% & 23.81\% & 0.5 \\
37.78 & 54 & 85\% & 85.4\% & 1.6 & 13.61\% & 2.96\% & 13.93\% & 4\% & 19.86\% & - & 0.49\% & 4.87\% & 25.07\% & 0.4 \\
38.23 & 50 & 84.3\% & 85.7\% & 1.5 & 14.14\% & 2.96\% & 14.45\% & 4\% & 19.97\% & - & 0.49\% & 4.87\% & 25.45\% & 0.4 \\
38.7 & 67 & 84\% & 85.7\% & 2.0 & 12.22\% & 2.96\% & 12.57\% & 4\% & 19.84\% & - & 0.49\% & 4.87\% & 24.32\% & 0.5 \\
39.16 & 63 & 84\% & 85.7\% & 1.9 & 12.6\% & 2.96\% & 12.94\% & 4\% & 19.37\% & - & 0.49\% & 4.87\% & 24.14\% & 0.5 \\
39.64 & 50 & 83.2\% & 85.8\% & 1.5 & 14.14\% & 2.96\% & 14.45\% & 4\% & 19.24\% & - & 0.49\% & 4.87\% & 24.88\% & 0.4 \\
40.13 & 53 & 82.6\% & 86.1\% & 1.6 & 13.74\% & 2.96\% & 14.05\% & 4\% & 19.47\% & - & 0.49\% & 4.87\% & 24.83\% & 0.4 \\
40.62 & 30 & 82.9\% & 86.1\% & 0.9 & 18.26\% & 2.96\% & 18.5\% & 4\% & 19.83\% & - & 0.49\% & 4.87\% & 27.85\% & 0.2 \\
41.13 & 43 & 82\% & 86.1\% & 1.3 & 15.25\% & 2.96\% & 15.54\% & 4\% & 20.27\% & - & 0.49\% & 4.87\% & 26.31\% & 0.3 \\
41.65 & 44 & 81.4\% & 86.4\% & 1.3 & 15.08\% & 2.96\% & 15.36\% & 4\% & 20.35\% & - & 0.49\% & 4.87\% & 26.27\% & 0.3 \\
42.17 & 40 & 80.7\% & 86.2\% & 1.2 & 15.81\% & 2.96\% & 16.09\% & 4\% & 20.37\% & - & 0.49\% & 4.87\% & 26.72\% & 0.3 \\
42.71 & 49 & 79.7\% & 86.5\% & 1.4 & 14.29\% & 2.96\% & 14.59\% & 4\% & 20.76\% & - & 0.49\% & 4.87\% & 26.15\% & 0.4 \\
43.25 & 26 & 79.2\% & 86.7\% & 0.7 & 19.61\% & 2.96\% & 19.83\% & 4\% & 21.17\% & - & 0.49\% & 4.87\% & 29.69\% & 0.2 \\
43.81 & 21 & 78.8\% & 86.8\% & 0.6 & 21.82\% & 2.96\% & 22.02\% & 4\% & 21.51\% & - & 0.49\% & 4.87\% & 31.42\% & 0.2 \\
44.38 & 27 & 78.6\% & 86.9\% & 0.8 & 19.25\% & 2.96\% & 19.47\% & 4\% & 21.82\% & - & 0.49\% & 4.87\% & 29.92\% & 0.2 \\
44.95 & 19 & 78\% & 86.9\% & 0.5 & 22.94\% & 2.96\% & 23.13\% & 4\% & 22.11\% & - & 0.49\% & 4.87\% & 32.62\% & 0.2 \\
45.55 & 26 & 77.5\% & 87.2\% & 0.7 & 19.61\% & 2.96\% & 19.83\% & 4\% & 22.08\% & - & 0.49\% & 4.87\% & 30.35\% & 0.2 \\
46.15 & 29 & 76.6\% & 87.1\% & 0.8 & 18.57\% & 2.96\% & 18.8\% & 4\% & 21.99\% & - & 0.49\% & 4.87\% & 29.62\% & 0.2 \\
46.76 & 18 & 76.3\% & 87.3\% & 0.5 & 23.57\% & 2.96\% & 23.76\% & 4\% & 22.27\% & - & 0.49\% & 4.87\% & 33.17\% & 0.2 \\
47.39 & 18 & 76.1\% & 87.4\% & 0.5 & 23.57\% & 2.96\% & 23.76\% & 4\% & 22.61\% & - & 0.49\% & 4.87\% & 33.4\% & 0.2 \\
48.03 & 14 & 74.7\% & 87.6\% & 0.4 & 26.73\% & 2.96\% & 26.89\% & 4\% & 22.74\% & - & 0.49\% & 4.87\% & 35.78\% & 0.1 \\
48.69 & 15 & 74.4\% & 87.7\% & 0.4 & 25.82\% & 2.96\% & 25.99\% & 4\% & 22.84\% & - & 0.49\% & 4.87\% & 35.17\% & 0.1 \\
49.35 & 15 & 73.2\% & 87.6\% & 0.4 & 25.82\% & 2.96\% & 25.99\% & 4\% & 22.88\% & - & 0.49\% & 4.87\% & 35.2\% & 0.1 \\
50.04 & 6 & 72.5\% & 87.9\% & 0.2 & 40.82\% & 2.96\% & 40.93\% & 4\% & 23.18\% & - & 0.49\% & 4.87\% & 47.46\% & 0.1 \\
50.73 & 10 & 71.9\% & 88.2\% & 0.3 & 31.62\% & 2.96\% & 31.76\% & 4\% & 23.57\% & - & 0.49\% & 4.87\% & 40.06\% & 0.1 \\
51.44 & 7 & 71.1\% & 88.2\% & 0.2 & 37.8\% & 2.96\% & 37.91\% & 4\% & 23.45\% & - & 0.49\% & 4.87\% & 45.02\% & 0.1 \\
52.17 & 7 & 70\% & 88.3\% & 0.2 & 37.8\% & 2.96\% & 37.91\% & 4\% & 23.65\% & - & 0.49\% & 4.87\% & 45.13\% & 0.1 \\
52.92 & 6 & 69.2\% & 88.5\% & 0.2 & 40.82\% & 2.96\% & 40.93\% & 4\% & 24.39\% & - & 0.49\% & 4.87\% & 48.07\% & 0.1 \\
53.67 & 3 & 68.1\% & 88.6\% & 0.1 & 57.74\% & 2.96\% & 57.81\% & 4\% & 24.54\% & - & 0.49\% & 4.87\% & 63.12\% & 0 \\
54.45 & 3 & 67.4\% & 88.6\% & 0.1 & 57.74\% & 2.96\% & 57.81\% & 4\% & 24.48\% & - & 0.49\% & 4.87\% & 63.1\% & 0.1 \\
55.25 & 4 & 65.6\% & 88.7\% & 0.1 & 50\% & 2.96\% & 50.09\% & 4\% & 24.29\% & - & 0.49\% & 4.87\% & 56.02\% & 0.1 \\

\hline 
 \end{longtable}

\begin{longtable}[h]{|c|c|c|c|c|c|c|c|c|c|c|c|c|c|c|c|c|}
 \caption{$^{12}$C(n,d$_0$)$^{11}$B supplemental table. The table contains the incident neutron energy ($E_n$); counts from $^{12}$C(n,d$_0$)$^{11}$B (Counts: Peak); total counts in the fitting region (peak and background, Counts: Total); the charged particle detection efficiency correction for this channel (Corrections: Efficiency); the simulated neutron transmission correction for the flux distribution (Corrections: Transmission); the cross section for this channel (Cross section: $\sigma$ (mb)); the statistical uncertainties, broken into separate contributions from the counts from (n,d$_0$) at each energy (Statistical Uncertainty: $\sigma_{Peak}$), from the the background counts at each energy (Statistical Uncertainty: $\sigma_{Bkg}$), and from the (n,$\alpha$) counts at 14.1 MeV used for normalization (Statistical uncertainties: $\sigma_{N_{n,\alpha}}$); the total statistical uncertainties combining all statistical uncertainties (Statistical Uncertainties: $\sigma_{stat. total}$); and systematic uncertainties, which include the charged particle detection efficiency correction uncertainty (Systematic Uncertainties: $\sigma_{Eff.}$), the relative flux distribution uncertainty (Systematic Uncertainties:  $\sigma_{Flux}$), the fit/integration uncertainties (Systematic Uncertainties: $\sigma_{Fit}$), the incident neutron energy reconstruction uncertainty (systematic uncertainties: $\sigma_{NKE}$), and the normalization uncertainty based on the weighted average uncertainty of previous data (Systematic Uncertainties: $\sigma_{Norm.}$). The total uncertainty, which includes both statistical and systematic uncertainties, is also included, both as a percentage and in terms of millibarns.\label{n,d0}}\\
 \hline
$E_n$  & \multicolumn{2}{c|}{Counts} & \multicolumn{2}{c|}{Corrections} & Cross section & \multicolumn{4}{c|}{Statistical Uncertainties} &  \multicolumn{5}{c|}{Systematic Uncertainties} & Total Unc. & Error bar\\
\hline    
(MeV) & Peak & Total & Efficiency & Transmission & $\sigma$ (mb) & $\sigma_{Peak}$ &  $\sigma_{Bkg}$ &  $\sigma_{N_{n,\alpha}}$ & $\sigma_{stat. total}$ & $\sigma_{Eff.}$ & $\sigma_{Flux}$ & $\sigma_{Fit}$ & $\sigma_{NKE}$ & $\sigma_{Norm.}$ & $\sigma_{Total}$ & Error (mb)\\
 \hline
 \endfirsthead
 \hline
$E_n$  & \multicolumn{2}{c|}{Counts} & \multicolumn{2}{c|}{Corrections} & Cross section & \multicolumn{4}{c|}{Statistical Uncertainties} &  \multicolumn{5}{c|}{Systematic Uncertainties} & Total Unc. & Error bar\\ \hline
(MeV) & Peak & Total & Efficiency & Transmission & $\sigma$ (mb) & $\sigma_{Peak}$ &  $\sigma_{Bkg}$ &  $\sigma_{N_{n,\alpha}}$ & $\sigma_{stat. total}$ & $\sigma_{Eff.}$ & $\sigma_{Flux}$ & $\sigma_{Fit}$ & $\sigma_{NKE}$ & $\sigma_{Norm.}$ & $\sigma_{Total}$ & Error (mb)\\ \hline
 \endhead

 \hline
 \endfoot
15.87 & 379 & 753 & 99.8\% & 80.5\% & 23.6 & 5.1\% & 5.1\% & 2.96\% & 7.8\% & 4\% & 5\% & 4.1\% & 0.49\% & 4.87\% & 11.99\% & 2.8 \\
16.07 & 596 & 966 & 100\% & 80.3\% & 37.0 & 4.1\% & 3.2\% & 2.96\% & 6\% & 4\% & 5\% & 2.2\% & 0.49\% & 4.87\% & 10.34\% & 3.8 \\
16.29 & 683 & 1067 & 99.8\% & 81\% & 41.9 & 3.8\% & 2.9\% & 2.96\% & 5.6\% & 4\% & 5\% & 2.4\% & 0.49\% & 4.87\% & 10.16\% & 4.3 \\
16.5 & 699 & 1119 & 99.9\% & 81.5\% & 42.1 & 3.8\% & 2.9\% & 2.96\% & 5.6\% & 4\% & 5\% & 2.4\% & 0.49\% & 4.87\% & 10.17\% & 4.3 \\
16.73 & 675 & 1077 & 99.9\% & 81.7\% & 40.1 & 3.8\% & 3\% & 2.96\% & 5.7\% & 4\% & 5\% & 2.2\% & 0.49\% & 4.87\% & 10.17\% & 4.1 \\
16.95 & 706 & 1122 & 99.9\% & 81.8\% & 41.9 & 3.8\% & 2.9\% & 2.96\% & 5.6\% & 4\% & 5\% & 2.3\% & 0.49\% & 4.87\% & 10.13\% & 4.2 \\
17.18 & 723 & 1043 & 99.8\% & 82.2\% & 43.1 & 3.7\% & 2.5\% & 2.96\% & 5.4\% & 4\% & 5\% & 1.9\% & 0.49\% & 4.87\% & 9.93\% & 4.3 \\
17.42 & 691 & 1025 & 99.6\% & 82.4\% & 40.8 & 3.8\% & 2.6\% & 2.96\% & 5.5\% & 4\% & 5.6\% & 2\% & 0.49\% & 4.87\% & 10.31\% & 4.2 \\
17.66 & 803 & 1134 & 99.5\% & 82\% & 47.2 & 3.5\% & 2.3\% & 2.96\% & 5.1\% & 4\% & 6.7\% & 2.1\% & 0.49\% & 4.87\% & 10.8\% & 5.1 \\
17.9 & 739 & 1073 & 99.2\% & 81.8\% & 43.4 & 3.7\% & 2.5\% & 2.96\% & 5.3\% & 4\% & 7.6\% & 2\% & 0.49\% & 4.87\% & 11.43\% & 5 \\
18.15 & 765 & 1063 & 99\% & 81.7\% & 44.7 & 3.6\% & 2.3\% & 2.96\% & 5.2\% & 4\% & 7.5\% & 2\% & 0.49\% & 4.87\% & 11.32\% & 5.1 \\
18.41 & 827 & 1113 & 98.8\% & 81.6\% & 48.1 & 3.5\% & 2\% & 2.96\% & 5\% & 4\% & 7.3\% & 1.8\% & 0.49\% & 4.87\% & 11.06\% & 5.3 \\
18.67 & 972 & 1245 & 98.5\% & 81.7\% & 56.5 & 3.2\% & 1.7\% & 2.96\% & 4.7\% & 4\% & 6.8\% & 1.4\% & 0.49\% & 4.87\% & 10.56\% & 6 \\
18.94 & 871 & 1101 & 98\% & 81.7\% & 50.9 & 3.4\% & 1.7\% & 2.96\% & 4.8\% & 4\% & 6.3\% & 1.8\% & 0.49\% & 4.87\% & 10.33\% & 5.3 \\
19.21 & 774 & 971 & 97.8\% & 81.1\% & 45.8 & 3.6\% & 1.8\% & 2.96\% & 5\% & 4\% & 5.6\% & 1.8\% & 0.49\% & 4.87\% & 10.05\% & 4.6 \\
19.49 & 697 & 891 & 97.2\% & 80.9\% & 41.3 & 3.8\% & 2\% & 2.96\% & 5.2\% & 4\% & 5.5\% & 1.7\% & 0.49\% & 4.87\% & 10.03\% & 4.1 \\
19.78 & 703 & 880 & 96.7\% & 80.9\% & 41.5 & 3.8\% & 1.9\% & 2.96\% & 5.2\% & 4\% & 5.5\% & 1.9\% & 0.49\% & 4.87\% & 10.06\% & 4.2 \\
20.07 & 697 & 861 & 96.4\% & 81.1\% & 40.8 & 3.8\% & 1.8\% & 2.96\% & 5.1\% & 4\% & 5.4\% & 2\% & 0.49\% & 4.87\% & 10.02\% & 4.1 \\
20.37 & 693 & 845 & 96\% & 81.5\% & 40.1 & 3.8\% & 1.8\% & 2.96\% & 5.1\% & 4\% & 5.5\% & 1.8\% & 0.49\% & 4.87\% & 10.04\% & 4 \\
20.67 & 652 & 806 & 95.2\% & 81.9\% & 37.2 & 3.9\% & 1.9\% & 2.96\% & 5.3\% & 4\% & 5.8\% & 2\% & 0.49\% & 4.87\% & 10.33\% & 3.8 \\
20.98 & 576 & 709 & 95\% & 81.9\% & 33.0 & 4.2\% & 2\% & 2.96\% & 5.5\% & 4\% & 5.3\% & 2.2\% & 0.49\% & 4.87\% & 10.19\% & 3.4 \\
21.3 & 533 & 685 & 93.9\% & 82.1\% & 30.5 & 4.3\% & 2.3\% & 2.96\% & 5.7\% & 4\% & 5.4\% & 2.2\% & 0.49\% & 4.87\% & 10.39\% & 3.2 \\
21.63 & 565 & 691 & 93\% & 82.2\% & 32.0 & 4.2\% & 2\% & 2.96\% & 5.5\% & 4\% & 6\% & 2.1\% & 0.49\% & 4.87\% & 10.56\% & 3.4 \\
21.96 & 574 & 711 & 92.3\% & 82.3\% & 32.3 & 4.2\% & 2\% & 2.96\% & 5.5\% & 4\% & 6.1\% & 2.3\% & 0.49\% & 4.87\% & 10.68\% & 3.5 \\
22.31 & 571 & 728 & 92\% & 82.1\% & 32.1 & 4.2\% & 2.2\% & 2.96\% & 5.6\% & 4\% & 6\% & 2.2\% & 0.49\% & 4.87\% & 10.65\% & 3.4 \\
22.66 & 650 & 827 & 91.3\% & 82.2\% & 36.6 & 3.9\% & 2\% & 2.96\% & 5.3\% & 4\% & 6\% & 2.1\% & 0.49\% & 4.87\% & 10.47\% & 3.8 \\
23.01 & 612 & 775 & 90.9\% & 82\% & 34.0 & 4\% & 2.1\% & 2.96\% & 5.4\% & 4\% & 6.5\% & 2.2\% & 0.49\% & 4.87\% & 10.86\% & 3.7 \\
23.38 & 513 & 706 & 89.8\% & 82.1\% & 28.1 & 4.4\% & 2.7\% & 2.96\% & 6\% & 4\% & 7.4\% & 3\% & 0.49\% & 4.87\% & 11.87\% & 3.3 \\
23.76 & 498 & 692 & 87.9\% & 81.9\% & 27.2 & 4.5\% & 2.8\% & 2.96\% & 6.1\% & 4\% & 8.5\% & 3.1\% & 0.49\% & 4.87\% & 12.6\% & 3.4 \\
24.14 & 448 & 617 & 87.5\% & 82\% & 24.1 & 4.7\% & 2.9\% & 2.96\% & 6.3\% & 4\% & 9.1\% & 3.3\% & 0.49\% & 4.87\% & 13.2\% & 3.2 \\
24.54 & 411 & 583 & 86.5\% & 82.1\% & 22.1 & 4.9\% & 3.2\% & 2.96\% & 6.6\% & 4\% & 9.3\% & 3.4\% &0.49\% & 4.87\% & 13.48\% & 3 \\
24.94 & 415 & 592 & 85\% & 82.2\% & 22.5 & 4.9\% & 3.2\% & 2.96\% & 6.6\% & 4\% & 9.3\% & 3.8\% & 0.49\% & 4.87\% & 13.61\% & 3.1 \\
25.36 & 445 & 597 & 84.1\% & 82.5\% & 24.0 & 4.7\% & 2.8\% & 2.96\% & 6.2\% & 4\% & 9.7\% & 2.7\% & 0.49\% & 4.87\% & 13.46\% & 3.2 \\
25.79 & 481 & 623 & 83.3\% & 82.6\% & 26.0 & 4.6\% & 2.5\% & 2.96\% & 6\% & 4\% & 10\% & 2.9\% & 0.49\% & 4.87\% & 13.57\% & 3.5 \\
26.22 & 407 & 596 & 81.8\% & 82.6\% & 22.2 & 5\% & 3.4\% & 2.96\% & 6.7\% & 4\% & 10.2\% & 3.7\% & 0.49\% & 4.87\% & 14.24\% & 3.2 \\
26.67 & 332 & 521 & 79.3\% & 83.1\% & 18.3 & 5.5\% & 4.1\% & 2.96\% & 7.5\% & 4\% & 10.8\% & 3.7\% & 0.49\% & 4.87\% & 15.03\% & 2.7 \\
27.13 & 285 & 463 & 77.8\% & 83.3\% & 15.4 & 5.9\% & 4.7\% & 2.96\% & 8.1\% & 4\% & 12.3\% & 5.3\% & 0.49\% & 4.87\% & 16.94\% & 2.6 \\
27.6 & 350 & 499 & 76.7\% & 83.4\% & 19.2 & 5.3\% & 3.5\% & 2.96\% & 7\% & 4\% & 12.3\% & 3\% & 0.49\% & 4.87\% & 15.81\% & 3 \\
28.09 & 303 & 471 & 75.6\% & 83.3\% & 16.9 & 5.7\% & 4.3\% & 2.96\% & 7.7\% & 4\% & 12.5\% & 5.7\% & 0.49\% & 4.87\% & 17.06\% & 2.9 \\
28.59 & 305 & 458 & 73.1\% & 83.6\% & 17.0 & 5.7\% & 4\% & 2.96\% & 7.6\% & 4\% & 13.8\% & 4.4\% & 0.49\% & 4.87\% & 17.55\% & 3 \\
29.1 & 299 & 448 & 70.9\% & 83.6\% & 16.8 & 5.8\% & 4.1\% & 2.96\% & 7.7\% & 4\% & 14\% & 4.9\% & 0.49\% & 4.87\% & 17.89\% & 3 \\
29.62 & 265 & 426 & 70\% & 83.7\% & 14.9 & 6.1\% & 4.8\% & 2.96\% & 8.3\% & 4\% & 14\% & 3.2\% & 0.49\% & 4.87\% & 17.77\% & 2.6 \\
30.16 & 306 & 441 & 66.7\% & 83.7\% & 17.9 & 5.7\% & 3.8\% & 2.96\% & 7.5\% & 4\% & 13.8\% & 3.3\% & 0.49\% & 4.87\% & 17.24\% & 3.1 \\
30.72 & 289 & 464 & 63.5\% & 84\% & 17.5 & 5.9\% & 4.6\% & 2.96\% & 8\% & 4\% & 13.6\% & 3.8\% & 0.49\% & 4.87\% & 17.5\% & 3.1 \\
31.29 & 324 & 517 & 61.5\% & 84.2\% & 19.8 & 5.6\% & 4.3\% & 2.96\% & 7.6\% & 4\% & 13.9\% & 3\% & 0.49\% & 4.87\% & 17.4\% & 3.4 \\
31.88 & 242 & 433 & 57.9\% & 84.3\% & 15.4 & 6.4\% & 5.7\% & 2.96\% & 9.1\% & 4\% & 14.1\% & 3.9\% & 0.49\% & 4.87\% & 18.41\% & 2.8 \\
32.48 & 244 & 417 & 55.5\% & 84.3\% & 15.8 & 6.4\% & 5.4\% & 2.96\% & 8.9\% & 4\% & 14.9\% & 3.8\% & 0.49\% & 4.87\% & 18.85\% & 3 \\
33.1 & 190 & 313 & 53.5\% & 84.6\% & 12.5 & 7.3\% & 5.8\% & 2.96\% & 9.8\% & 4\% & 15.3\% & 4.5\% & 0.49\% & 4.87\% & 19.74\% & 2.5 \\
33.75 & 229 & 409 & 50.8\% & 84.7\% & 15.5 & 6.6\% & 5.8\% & 2.96\% & 9.3\% & 4\% & 15.6\% & 3.9\% & 0.49\% & 4.87\% & 19.67\% & 3.1 \\
34.41 & 202 & 393 & 48.4\% & 85\% & 13.9 & 7\% & 6.9\% & 2.96\% & 10.3\% & 4\% & 16.5\% & 5.7\% & 0.49\% & 4.87\% & 21.27\% & 3 \\
35.08 & 198 & 365 & 46.8\% & 84.9\% & 13.7 & 7.1\% & 6.5\% & 2.96\% & 10.1\% & 4\% & 17.5\% & 4.5\% & 0.49\% & 4.87\% & 21.68\% & 3 \\
35.79 & 170 & 300 & 45.5\% & 85.2\% & 11.9 & 7.7\% & 6.7\% & 2.96\% & 10.6\% & 4\% & 17.7\% & 7.1\% & 0.49\% & 4.87\% & 22.77\% & 2.7 \\
36.51 & 169 & 318 & 43.5\% & 85.4\% & 12.1 & 7.7\% & 7.2\% & 2.96\% & 10.9\% & 4\% & 18\% & 5.1\% & 0.49\% & 4.87\% & 22.6\% & 2.7 \\
37.25 & 170 & 312 & 41.8\% & 85.4\% & 12.1 & 7.7\% & 7\% & 2.96\% & 10.8\% & 4\% & 19.2\% & 6.3\% & 0.49\% & 4.87\% & 23.78\% & 2.9 \\
38.02 & 143 & 296 & 40.2\% & 85.5\% & 10.3 & 8.4\% & 8.6\% & 2.96\% & 12.4\% & 4\% & 19.9\% & 10\% & 0.49\% & 4.87\% & 26.29\% & 2.7 \\
38.81 & 84 & 254 & 39.9\% & 85.7\% & 6.0 & 10.9\% & 15.5\% & 2.96\% & 19.2\% & 4\% & 19.7\% & 9.2\% & 0.49\% & 4.87\% & 29.71\% & 1.8 \\
39.63 & 66 & 149 & 38.2\% & 85.8\% & 4.9 & 12.3\% & 13.8\% & 2.96\% & 18.8\% & 4\% & 19.2\% & 12.8\% & 0.49\% & 4.87\% & 30.46\% & 1.5 \\
40.47 & 91 & 278 & 36.7\% & 86.1\% & 6.8 & 10.5\% & 15.1\% & 2.96\% & 18.7\% & 4\% & 19.7\% & 9.1\% & 0.49\% & 4.87\% & 29.34\% & 2 \\
41.35 & 78 & 206 & 35.1\% & 86.2\% & 6.0 & 11.3\% & 14.4\% & 2.96\% & 18.5\% & 4\% & 20.3\% & 6.7\% & 0.49\% & 4.87\% & 29.01\% & 1.7 \\
42.25 & 80 & 200 & 34.1\% & 86.2\% & 6.2 & 11.2\% & 13.6\% & 2.96\% & 17.9\% & 4\% & 20.4\% & 14.2\% & 0.49\% & 4.87\% & 31.29\% & 2 \\
43.18 & 72 & 226 & 33.1\% & 86.7\% & 5.6 & 11.8\% & 17.1\% & 2.96\% & 21\% & 4\% & 21.1\% & 7.2\% & 0.49\% & 4.87\% & 31.3\% & 1.8 \\
44.14 & 62 & 163 & 30.8\% & 86.8\% & 5.1 & 12.7\% & 16.3\% & 2.96\% & 20.8\% & 4\% & 21.7\% & 11.6\% & 0.49\% & 4.87\% & 32.87\% & 1.7 \\
45.14 & 87 & 195 & 30\% & 87\% & 7.2 & 10.7\% & 11.9\% & 2.96\% & 16.3\% & 4\% & 22.1\% & 9.2\% & 0.49\% & 4.87\% & 29.64\% & 2.1 \\
\hline 
 \end{longtable}
 
\captionsetup{width=.9\textwidth}
\begin{table}[]
\caption{\label{tab:n_p0_xs} $^{12}$C(n,p$_0$)$^{12}$B supplemental table. The table contains the incident neutron energy ($E_n$); counts from $^{12}$C(n,p$_0$)$^{12}$B (Counts: Peak); total counts in the fitting region (peak and background, Counts: Total); the charged particle detection efficiency correction for this channel (Corrections: Efficiency); the simulated neutron transmission correction for the flux distribution (Corrections: Transmission); the cross section for this channel (Cross section: $\sigma$ (mb)); the statistical uncertainties, broken into separate contributions from the counts from (n,p$_0$) at each energy (Statistical Uncertainty: $\sigma_{Peak}$), from the the background counts at each energy (Statistical Uncertainty: $\sigma_{Bkg}$), and from the (n,$\alpha$) counts at 14.1 MeV used for normalization (Statistical uncertainties: $\sigma_{N_{n,\alpha}}$); the total statistical uncertainties combining all statistical uncertainties (Statistical Uncertainties: $\sigma_{stat. total}$); and systematic uncertainties, which include the charged particle detection efficiency correction uncertainty (Systematic Uncertainties: $\sigma_{Eff.}$), the relative flux distribution uncertainty (Systematic Uncertainties:  $\sigma_{Flux}$), the fit/integration uncertainties (Systematic Uncertainties: $\sigma_{Fit}$), the incident neutron energy reconstruction uncertainty (systematic uncertainties: $\sigma_{NKE}$), and the normalization uncertainty based on the weighted average uncertainty of previous data (Systematic Uncertainties: $\sigma_{Norm.}$). The total uncertainty, which includes both statistical and systematic uncertainties, is also included, both as a percentage and in terms of millibarns. }
    \centering
    \begin{tabular}{|c|c|c|c|c|c|c|c|c|c|c|c|c|c|c|c|c|}
    \hline
$E_n$ & \multicolumn{2}{c|}{Counts} & \multicolumn{2}{c|}{Corrections} & Cross section & \multicolumn{4}{c|}{Statistical Uncertainties} &  \multicolumn{5}{c|}{Systematic Uncertainties} & Total Unc. & Error bar\\
\hline    
(MeV) & Peak & Total & Efficiency & Transmission & $\sigma$ (mb) & $\sigma_{Peak}$ &  $\sigma_{Bkg}$ &  $\sigma_{N_{n,\alpha}}$ & $\sigma_{stat. total}$ & $\sigma_{Eff.}$ & $\sigma_{Flux}$ & $\sigma_{Fit}$ & $\sigma_{NKE}$ & $\sigma_{Norm.}$ & $\sigma_{Total}$ & Error (mb)\\
\hline
15.71 & 399 & 1268 & 98.4\% & 80.6\% & 16.7 & 5\% & 8.9\% & 2.96\% & 10.6\% & 4\% & 5\% & 4.5\% & 0.49\% & 4.87\% & 13.16\% & 2.2 \\
16.02 & 485 & 1231 & 97.1\% & 80.2\% & 20.6 & 4.5\% & 7.2\% & 2.96\% & 9\% & 4\% & 5\% & 4.3\% & 0.49\% & 4.87\% & 12.02\% & 2.5 \\
16.34 & 467 & 1054 & 96.4\% & 81.2\% & 19.6 & 4.6\% & 7\% & 2.96\% & 8.9\% & 4\% & 5\% & 3.8\% & 0.49\% & 4.87\% & 11.68\% & 2.3 \\
16.67 & 327 & 776 & 96.9\% & 81.6\% & 13.3 & 5.5\% & 8.5\% & 2.96\% & 10.6\% & 4\% & 5\% & 3.8\% & 0.49\% & 4.87\% & 12.67\% & 1.7 \\
17.01 & 360 & 657 & 95.2\% & 81.9\% & 14.9 & 5.3\% & 7.1\% & 2.96\% & 9.3\% & 4\% & 5\% & 2.9\% & 0.49\% & 4.87\% & 11.5\% & 1.7 \\
17.36 & 505 & 880 & 94.8\% & 82.2\% & 20.8 & 4.5\% & 5.9\% & 2.96\% & 7.9\% & 4\% & 5.3\% & 3.6\% & 0.49\% & 4.87\% & 11.12\% & 2.3 \\
17.72 & 443 & 761 & 93.6\% & 82\% & 18.2 & 4.8\% & 6.2\% & 2.96\% & 8.4\% & 4\% & 6.9\% & 3.5\% & 0.49\% & 4.87\% & 12.16\% & 2.2 \\
18.09 & 315 & 505 & 92.2\% & 81.7\% & 13.1 & 5.6\% & 7.1\% & 2.96\% & 9.6\% & 4\% & 7.5\% & 3.1\% & 0.49\% & 4.87\% & 12.86\% & 1.7 \\
18.48 & 314 & 485 & 90.9\% & 81.7\% & 13.1 & 5.6\% & 7\% & 2.96\% & 9.5\% & 4\% & 7.2\% & 3.1\% & 0.49\% & 4.87\% & 12.61\% & 1.7 \\
18.87 & 246 & 426 & 89\% & 81.7\% & 10.5 & 6.4\% & 8.4\% & 2.96\% & 11\% & 4\% & 6.4\% & 3.8\% & 0.49\% & 4.87\% & 13.21\% & 1.4 \\
19.28 & 275 & 481 & 89.2\% & 81\% & 11.8 & 6\% & 8\% & 2.96\% & 10.4\% & 4\% & 5.5\% & 3.6\% & 0.49\% & 4.87\% & 12.49\% & 1.5 \\
19.71 & 220 & 377 & 86\% & 80.7\% & 9.7 & 6.7\% & 8.8\% & 2.96\% & 11.5\% & 4\% & 5.5\% & 3.5\% & 0.49\% & 4.87\% & 12.99\% & 1.3 \\
20.14 & 235 & 401 & 84.4\% & 81.1\% & 10.4 & 6.5\% & 8.5\% & 2.96\% & 11.1\% & 4\% & 5.4\% & 4\% & 0.49\% & 4.87\% & 12.89\% & 1.3 \\
20.6 & 201 & 369 & 83.2\% & 81.8\% & 8.7 & 7.1\% & 9.6\% & 2.96\% & 12.3\% & 4\% & 5.8\% & 3.9\% & 0.49\% & 4.87\% & 13.73\% & 1.2 \\
21.06 & 142 & 309 & 81.4\% & 81.9\% & 6.3 & 8.4\% & 12.4\% & 2.96\% & 15.2\% & 4\% & 5.2\% & 7\% & 0.49\% & 4.87\% & 16.68\% & 1 \\
21.55 & 133 & 313 & 79.4\% & 81.9\% & 5.9 & 8.7\% & 13.3\% & 2.96\% & 16.2\% & 4\% & 5.9\% & 6\% & 0.49\% & 4.87\% & 17.22\% & 1 \\
22.05 & 127 & 329 & 76.6\% & 82.4\% & 5.7 & 8.9\% & 14.3\% & 2.96\% & 17.1\% & 4\% & 6.1\% & 6.3\% & 0.49\% & 4.87\% & 18.15\% & 1 \\
22.57 & 137 & 301 & 73.4\% & 82\% & 6.4 & 8.5\% & 12.7\% & 2.96\% & 15.6\% & 4\% & 5.9\% & 7.3\% & 0.49\% & 4.87\% & 17.23\% & 1.1 \\
23.11 & 147 & 314 & 70.7\% & 82.1\% & 6.9 & 8.2\% & 12\% & 2.96\% & 14.9\% & 4\% & 6.7\% & 4.9\% & 0.49\% & 4.87\% & 16.19\% & 1.1 \\
23.66 & 189 & 351 & 68.1\% & 81.9\% & 8.9 & 7.3\% & 9.9\% & 2.96\% & 12.7\% & 4\% & 8.3\% & 3.8\% & 0.49\% & 4.87\% & 15.15\% & 1.3 \\
24.24 & 163 & 338 & 65.3\% & 81.8\% & 7.7 & 7.8\% & 11.3\% & 2.96\% & 14\% & 4\% & 9.3\% & 5\% & 0.49\% & 4.87\% & 16.93\% & 1.3 \\
24.84 & 128 & 314 & 59.9\% & 82.4\% & 6.5 & 8.8\% & 13.8\% & 2.96\% & 16.7\% & 4\% & 9.3\% & 5.9\% & 0.49\% & 4.87\% & 18.97\% & 1.2 \\
25.47 & 78 & 268 & 57.2\% & 82.4\% & 4.1 & 11.4\% & 21.1\% & 2.96\% & 24.2\% & 4\% & 9.8\% & 14.9\% & 0.49\% & 4.87\% & 28.5\% & 1.2 \\
26.11 & 89 & 259 & 53.2\% & 82.6\% & 5.0 & 10.6\% & 18\% & 2.96\% & 21.1\% & 4\% & 10.1\% & 6.5\% & 0.49\% & 4.87\% & 22.74\% & 1.1 \\
26.79 & 66 & 224 & 49.5\% & 83.2\% & 3.8 & 12.3\% & 22.5\% & 2.96\% & 25.8\% & 4\% & 11.1\% & 7.8\% & 0.49\% & 4.87\% & 27.2\% & 1 \\
27.49 & 59 & 245 & 45\% & 83.4\% & 3.7 & 13\% & 26.5\% & 2.96\% & 29.6\% & 4\% & 12.4\% & 11.8\% & 0.49\% & 4.87\% & 32.27\% & 1.2 \\
\hline
    \end{tabular}
\end{table}

\begin{table}[]
\caption{\label{tab:n_p1_xs} $^{12}$C(n,p$_1$)$^{12}$B supplemental table. The table contains the incident neutron energy ($E_n$); counts from $^{12}$C(n,p$_1$)$^{12}$B (Counts: Peak); total counts in the fitting region (peak and background, Counts: Total); the charged particle detection efficiency correction for this channel (Corrections: Efficiency); the simulated neutron transmission correction for the flux distribution (Corrections: Transmission); the cross section for this channel (Cross section: $\sigma$ (mb)); the statistical uncertainties, broken into separate contributions from the counts from (n,p$_1$) at each energy (Statistical Uncertainty: $\sigma_{Peak}$), from the the background counts at each energy (Statistical Uncertainty: $\sigma_{Bkg}$), and from the (n,$\alpha$) counts at 14.1 MeV used for normalization (Statistical uncertainties: $\sigma_{N_{n,\alpha}}$); the total statistical uncertainties combining all statistical uncertainties (Statistical Uncertainties: $\sigma_{stat. total}$); and systematic uncertainties, which include the charged particle detection efficiency correction uncertainty (Systematic Uncertainties: $\sigma_{Eff.}$), the relative flux distribution uncertainty (Systematic Uncertainties:  $\sigma_{Flux}$), the fit/integration uncertainties (Systematic Uncertainties: $\sigma_{Fit}$), the incident neutron energy reconstruction uncertainty (systematic uncertainties: $\sigma_{NKE}$), and the normalization uncertainty based on the weighted average uncertainty of previous data (Systematic Uncertainties: $\sigma_{Norm.}$). The total uncertainty, which includes both statistical and systematic uncertainties, is also included, both as a percentage and in terms of millibarns.}
    \centering
    \begin{tabular}{|c|c|c|c|c|c|c|c|c|c|c|c|c|c|c|c|c|}
    \hline
$E_n$  & \multicolumn{2}{c|}{Counts} & \multicolumn{2}{c|}{Corrections} & Cross section & \multicolumn{4}{c|}{Statistical Uncertainties} &  \multicolumn{5}{c|}{Systematic Uncertainties} & Total Unc. & Error bar\\
\hline    
(MeV) & Peak & Total & Efficiency & Transmission & $\sigma$ (mb) & $\sigma_{Peak}$ &  $\sigma_{Bkg}$ &  $\sigma_{N_{n,\alpha}}$ & $\sigma_{stat. total}$ & $\sigma_{Eff.}$ & $\sigma_{Flux}$ & $\sigma_{Fit}$ & $\sigma_{NKE}$ & $\sigma_{Norm.}$ & $\sigma_{Total}$ & Error (mb)\\
\hline
16.5 & 134 & 745 & 98.5\% & 81.5\% & 8.2 & 8.6\% & 18.5\% & 2.9\% & 20.6\% & 4\% & 5\% & 11.2\% & 0.49\% & 4.87\% & 24.78\% & 2 \\
16.73 & 133 & 488 & 98.5\% & 81.7\% & 8.0 & 8.7\% & 14.2\% & 2.9\% & 16.9\% & 4\% & 5\% & 7.2\% & 0.49\% & 4.87\% & 20.08\% & 1.6 \\
16.95 & 79 & 298 & 97.9\% & 81.8\% & 4.8 & 11.3\% & 18.7\% & 2.9\% & 22\% & 4\% & 5\% & 10.1\% & 0.49\% & 4.87\% & 25.57\% & 1.2 \\
17.18 & 161 & 587 & 97.6\% & 82.2\% & 9.8 & 7.9\% & 12.8\% & 2.9\% & 15.3\% & 4\% & 5\% & 6.9\% & 0.49\% & 4.87\% & 18.66\% & 1.8 \\
17.42 & 80 & 230 & 95.7\% & 82.4\% & 4.9 & 11.2\% & 15.3\% & 2.9\% & 19.2\% & 4\% & 5.6\% & 25\% & 0.49\% & 4.87\% & 32.64\% & 1.6 \\
17.66 & 247 & 787 & 96.6\% & 82\% & 14.9 & 6.4\% & 9.4\% & 2.9\% & 11.7\% & 4\% & 6.7\% & 6.1\% & 0.49\% & 4.87\% & 16.14\% & 2.4 \\
17.9 & 187 & 699 & 96.1\% & 81.8\% & 11.3 & 7.3\% & 12.1\% & 2.9\% & 14.4\% & 4\% & 7.6\% & 6.4\% & 0.49\% & 4.87\% & 18.59\% & 2.1 \\
18.15 & 154 & 374 & 95.4\% & 81.7\% & 9.3 & 8.1\% & 9.6\% & 2.9\% & 12.9\% & 4\% & 7.5\% & 6.3\% & 0.49\% & 4.87\% & 17.38\% & 1.6 \\
18.41 & 118 & 344 & 95.3\% & 81.6\% & 7.1 & 9.2\% & 12.7\% & 2.9\% & 15.9\% & 4\% & 7.3\% & 5.7\% & 0.49\% & 4.87\% & 19.48\% & 1.4 \\
18.67 & 116 & 351 & 93.6\% & 81.7\% & 7.1 & 9.3\% & 13.2\% & 2.9\% & 16.4\% & 4\% & 6.8\% & 13.8\% & 0.49\% & 4.87\% & 23.38\% & 1.7 \\
18.94 & 206 & 673 & 93.3\% & 81.7\% & 12.6 & 7\% & 10.5\% & 2.9\% & 13\% & 4\% & 6.3\% & 10\% & 0.49\% & 4.87\% & 18.63\% & 2.4 \\
19.21 & 126 & 261 & 93\% & 81.1\% & 7.9 & 8.9\% & 9.2\% & 2.9\% & 13.1\% & 4\% & 5.6\% & 7.6\% & 0.49\% & 4.87\% & 17.34\% & 1.4 \\
19.49 & 156 & 296 & 92.7\% & 80.9\% & 9.7 & 8\% & 7.6\% & 2.9\% & 11.4\% & 4\% & 5.5\% & 4.8\% & 0.49\% & 4.87\% & 14.96\% & 1.4 \\
19.78 & 215 & 483 & 91.2\% & 80.9\% & 13.4 & 6.8\% & 7.6\% & 2.9\% & 10.6\% & 4\% & 5.5\% & 4.3\% & 0.49\% & 4.87\% & 14.2\% & 1.9 \\
20.07 & 221 & 569 & 88.6\% & 81.1\% & 14.0 & 6.7\% & 8.5\% & 2.9\% & 11.2\% & 4\% & 5.4\% & 5.5\% & 0.49\% & 4.87\% & 14.97\% & 2.1 \\
20.37 & 231 & 412 & 90.4\% & 81.5\% & 14.2 & 6.6\% & 5.8\% & 2.9\% & 9.2\% & 4\% & 5.5\% & 4.2\% & 0.49\% & 4.87\% & 13.17\% & 1.9 \\
20.67 & 201 & 346 & 87.6\% & 81.9\% & 12.4 & 7.1\% & 6\% & 2.9\% & 9.7\% & 4\% & 5.8\% & 3.9\% & 0.49\% & 4.87\% & 13.55\% & 1.7 \\
20.98 & 184 & 444 & 85.5\% & 81.9\% & 11.7 & 7.4\% & 8.8\% & 2.9\% & 11.8\% & 4\% & 5.3\% & 5.9\% & 0.49\% & 4.87\% & 15.56\% & 1.8 \\
21.3 & 112 & 207 & 84.9\% & 82.1\% & 7.1 & 9.5\% & 8.8\% & 2.9\% & 13.2\% & 4\% & 5.4\% & 11.3\% & 0.49\% & 4.87\% & 19.29\% & 1.4 \\
21.63 & 189 & 324 & 83.1\% & 82.2\% & 12.0 & 7.3\% & 6.1\% & 2.9\% & 10\% & 4\% & 6\% & 4.2\% & 0.49\% & 4.87\% & 13.88\% & 1.7 \\
21.96 & 170 & 212 & 82.2\% & 82.3\% & 10.8 & 7.7\% & 3.8\% & 2.9\% & 9\% & 4\% & 6.1\% & 10\% & 0.49\% & 4.87\% & 16.09\% & 1.7 \\
22.31 & 186 & 235 & 79\% & 82.1\% & 12.2 & 7.3\% & 3.7\% & 2.9\% & 8.7\% & 4\% & 6\% & 6.5\% & 0.49\% & 4.87\% & 13.93\% & 1.7 \\
22.66 & 181 & 250 & 79.9\% & 82.2\% & 11.7 & 7.4\% & 4.6\% & 2.9\% & 9.2\% & 4\% & 6\% & 6.2\% & 0.49\% & 4.87\% & 14.1\% & 1.6 \\
23.01 & 170 & 219 & 77.1\% & 82\% & 11.1 & 7.7\% & 4.1\% & 2.9\% & 9.2\% & 4\% & 6.5\% & 6.1\% & 0.49\% & 4.87\% & 14.29\% & 1.6 \\
23.38 & 167 & 284 & 75.6\% & 82.1\% & 10.9 & 7.7\% & 6.5\% & 2.9\% & 10.5\% & 4\% & 7.4\% & 12.3\% & 0.49\% & 4.87\% & 18.9\% & 2.1 \\
23.76 & 156 & 253 & 73\% & 81.9\% & 10.3 & 8\% & 6.3\% & 2.9\% & 10.6\% & 4\% & 8.5\% & 9.6\% & 0.49\% & 4.87\% & 17.79\% & 1.8 \\
24.14 & 127 & 257 & 70.4\% & 82\% & 8.5 & 8.9\% & 9\% & 2.9\% & 13\% & 4\% & 9.1\% & 16.2\% & 0.49\% & 4.87\% & 23.49\% & 2 \\
24.54 & 140 & 324 & 68.5\% & 82.1\% & 9.5 & 8.5\% & 9.7\% & 2.9\% & 13.2\% & 4\% & 9.3\% & 9.7\% & 0.49\% & 4.87\% & 19.82\% & 1.9 \\
24.94 & 105 & 180 & 66.8\% & 82.2\% & 7.2 & 9.8\% & 8.3\% & 2.9\% & 13.1\% & 4\% & 9.3\% & 13.3\% & 0.49\% & 4.87\% & 21.8\% & 1.6 \\
25.36 & 62 & 107 & 62.5\% & 82.5\% & 4.5 & 12.7\% & 10.7\% & 2.9\% & 16.8\% & 4\% & 9.7\% & 14.2\% & 0.49\% & 4.87\% & 24.89\% & 1.1 \\
25.79 & 71 & 122 & 60\% & 82.6\% & 5.3 & 11.9\% & 10\% & 2.9\% & 15.8\% & 4\% & 10\% & 15.8\% & 0.49\% & 4.87\% & 25.27\% & 1.3 \\
26.22 & 89 & 283 & 57.3\% & 82.6\% & 6.9 & 10.6\% & 15.6\% & 2.9\% & 19.1\% & 4\% & 10.2\% & 27\% & 0.49\% & 4.87\% & 35.16\% & 2.4 \\
26.67 & 144 & 397 & 55.1\% & 83.1\% & 11.4 & 8.3\% & 11.1\% & 2.9\% & 14.2\% & 4\% & 10.8\% & 16\% & 0.49\% & 4.87\% & 24.75\% & 2.8 \\
27.13 & 133 & 361 & 53\% & 83.3\% & 10.6 & 8.7\% & 11.4\% & 2.9\% & 14.6\% & 4\% & 12.3\% & 15.9\% & 0.49\% & 4.87\% & 25.62\% & 2.7 \\
27.6 & 49 & 95 & 50.6\% & 83.4\% & 4.1 & 14.2\% & 13.6\% & 2.9\% & 19.9\% & 4\% & 12.3\% & 17.3\% & 0.49\% & 4.87\% & 29.77\% & 1.2 \\
\hline
    \end{tabular}
\end{table}